\documentclass{aa}

\pdfoutput=1
\usepackage{graphicx}
\usepackage{txfonts}
\usepackage{natbib}
\bibpunct{(}{)}{;}{a}{}{,} 

\begin{document}

\title{Digitization and astrometric calibration of Carte du Ciel photographic
  plates with {\it Gaia}~DR1}

\author{K.\ Lehtinen\inst{\ref{inst:FGI}} \and T.\ Prusti\inst{\ref{inst:esa}}
  \and J.\ de Bruijne\inst{\ref{inst:esa}} \and
  U.\ Lammers\inst{\ref{inst:esac}} \and
  C.F.\ Manara\inst{\ref{inst:esa},\ref{inst:eso}} \and
  J.-U.\ Ness\inst{\ref{inst:esa}} \and H.\ Siddiqui\inst{\ref{inst:esac}} \and
  T.\ Markkanen\inst{\ref{inst:hu}}\thanks{Deceased August 28, 2017} \and
  M.\ Poutanen\inst{\ref{inst:FGI}} \and
  K.\ Muinonen\inst{\ref{inst:FGI},\ref{inst:hu}} }

\institute{ Finnish Geospatial Research Institute, Geodeetinrinne 2, 02430
  Masala, Finland \email{kimmo.lehtinen@nls.fi} \label{inst:FGI} \and Science
  Support Office, Directorate of Science, European Space Research and Technology
  Centre (ESA/ESTEC), Keplerlaan 1, 2201AZ, Noordwijk, The
  Netherlands \label{inst:esa} \and European Space Astronomy Centre (ESA/ESAC),
  Camino bajo del Castillo, s/n, Urbanizacion Villafranca del Castillo,
  Villanueva de la Ca\~nada, 28692 Madrid, Spain \label{inst:esac} \and European
  Southern Observatory (ESO), Karl-Schwarzschild-Stra\ss e 2, 85748 Garching bei
  M\"unchen, Germany \label{inst:eso} \and Department of Physics, P.O.\ BOX 64,
  00014 University of Helsinki, Finland \label{inst:hu} }

\abstract 
{Carte du Ciel was a global international project at the end of the nineteenth and
  beginning of the twentieth century to map the sky to about magnitude 14 on
  photographic plates. The full project was never observationally completed and
  a large fraction of the observations made remain unanalyzed.}
{We want to study whether the astrometric and photometric accuracies obtained for the Carte
  du Ciel plates digitized with a commercial digital camera are high enough for
  scientific exploitation of the plates.}
{We use a digital camera Canon EOS~5Ds, with a 100\,mm macrolens for
  digitizing. We analyze six single-exposure plates and four triple-exposure
  plates from the Helsinki zone of Carte du Ciel
  ($+39\degr\le\delta\le+47\degr$). Each plate is digitized using four images,
  with a significant central area being covered twice for quality control
  purposes. The astrometric calibration of the digitized images is done with the
  data from the {\it Gaia} TGAS (Tycho-{\it Gaia} Astrometric Solution) of the
  first {\it Gaia} data release ({\it Gaia}~DR1), Tycho-2, HSOY (Hot Stuff for
  One Year), UCAC5 (USNO CCD Astrograph Catalog), and PMA catalogs.}
{ The best astrometric accuracy is obtained with the UCAC5 reference stars.  The
  astrometric accuracy for single-exposure plates is $\sigma(\alpha
  \cos(\delta))=0.16\arcsec$ and $\sigma(\delta)=0.15\arcsec$, expressed as
  a Gaussian deviation of the astrometric residuals.  For triple-exposure plates
  the astrometric accuracy is $\sigma(\alpha \cos(\delta))=0.12\arcsec$ and
  $\sigma(\delta)=0.13\arcsec$.  The 1-$\sigma$ uncertainty of photometric
  calibration is about 0.28\,mag and 0.24\,mag for single- and triple-exposure
  plates, respectively.  We detect the photographic adjacency (Kostinsky) effect
  in the triple-exposure plates.}
{We show that accuracies at least of the level of scanning machines can be
  achieved with a digital camera, without any corrections for possible
  distortions caused by our instrumental setup. This method can be used to
  rapidly and inexpensively digitize and calibrate old photographic plates
  enabling their scientific exploitation.}


\titlerunning{Carte du Ciel and {\it Gaia}}
\authorrunning{K.\ Lehtinen et al.}
\maketitle

\section{Introduction}

Carte du Ciel (hereafter CduC) was a massive international project, initiated in
the late nineteenth century, to make photographic charts of the whole sky
representing the relative positions of stars down to a photographic magnitude
limit of about 14\,mag.  The CduC plates proved to be expensive to photograph
and reproduce, thus the CduC survey was never fully finished.  However, a
related photographic survey called Astrographic catalog (or Astrographic Chart,
AC), with a lower limiting magnitude of about 11\,mag, was finished. The
scientific motivation of the AC was to determine the proper motions of stars and to
study the kinematics of stars in the solar neighborhood.  The Observatory of the
University of Helsinki completed the AC observations of its zone and measured a
total of 284663 star positions, which were published in eight volumes (in 12
books) during the years 1903-1937 (e.g.,\ \citet{Donner1894, Donner1902,
  Donner1908, Donner1929}).  The data of the AC survey have been used for the
Tycho-2 catalog \citep{Hog2000} proper motions by providing an early epoch
position, while the more recent position came from the measurements of the
Hipparcos satellite \citep{ESA1997} by ESA.  The availability of {\it Gaia} data
makes the CduC plates again interesting as the {\it Gaia} proper motions are of
such good quality that the {\it Gaia} reference frame can be translated beyond
a century without much loss of accuracy.  Therefore it is possible to establish
absolute astrometry for the CduC plates that is much less limited by the
uncertainty of proper motion values than in the case of the Tycho-2 catalog where
accuracy is limited only by precision achievable with the plates.

The optical system in the CduC telescopes is an aplanatic doublet objective,
which exhibits some astigmatism and field curvature \citep{Jones2000}.  The
photographic glass plates are most sensitive to blue light ($\sim$430\,nm), with
a limiting photographic magnitude of $\sim$14\,mag. The plates have a total size
of 16\,cm$\times$16\,cm, while the area covered by photographic emulsion is
13\,cm$\times$13\,cm.  To assist in the manual measuring of the star positions
and to monitor the possible emulsion shifts, the CduC plates have
photographically superposed r\'eseau grid lines with 5\,mm separation.  The
scale of the CduC plates is 1\,arcminute per millimetre, and the field of view
is 2\degr$\times$2\degr.  The plates of our study were taken at the Observatory
of the University of Helsinki, around the year 1900.  The Declination range of
Helsinki covered the range 39\degr-47\degr\ in 1900 equinox. In this paper we carry out
a study of a subset of ten plates in order to assess the achievable astrometric
accuracy.

The observations in Helsinki were made in a full overlap mode (the corner of a
plate is at the center of another plate).  The CduC plates along odd
Declinations were exposed three times. Each exposure time was 30 minutes, and
the pointing of the telescope was moved by about 10\arcsec\ between the
exposures. Thus, there are three images of each star, forming an equilateral
triangle, an asterism. When a star is brighter than $B\approx10-12$\,mag the
star images in the asterism start to merge together, and consequently an
adjacency photographic effect, the so-called Kostinsky effect \citep{Kostinski1907,
  Ross1921}, may arise. The Kostinsky effect increases the relative distances of
the stellar images in an asterism as a function of the flux of the star.  For
CduC triple images, \citet{Dick1993} and \citet{OrtizGil1998} have found the
Kostinsky effect, while \citet{Geffert1996} and \citet{BustosFierro2003} have
not found it.

Previously, the CduC plates have been digitized with microdensitometers
\citep{Dick1993, Geffert1996, Rapaport2006}, CCD (Charge Coupled Device) cameras
\citep{BustosFierro2003}, and flatbed scanners \citep{Vicente2007,
  Vicente2010}. Microdensitometers are now obsolete (too slow and expensive),
while scientific CCD cameras do not have enough pixels to image plates unless
one takes a mosaic with a large number of images.  Commercial scanners can
digitize large plates with good resolution in a reasonably short time, but their
problem is a large and non-constant distortion introduced by the scanner itself
\citep{Vicente2007, Vicente2010}.  However, for a digital camera and lens
combination, any distortion is constant, and once determined the distortion can
be removed from images.

While digitized Carte du Ciel plates have been previously mainly used for proper
motion studies, this is not meaningful in the {\it Gaia} era. Our main science
goal is to find binary stars with periodicity in the range of several decades to
about a century.  With {\it Gaia} astrometry it is possible to predict precisely
where we expect {\it Gaia} stars to be on the Carte du Ciel plates at their
observing epochs. Most are at the expected positions allowing astrometric
absolute calibration of the plates. The science is in the cases where the star
is not at the expected position. The astrophysical reason for this can be
binarity, which is not resolved by {\it Gaia}. For long period unresolved
binaries the {\it Gaia} proper motion is a combination of the system proper
motion and a part of the orbital motion. Thus positional mismatch may be an
indication of a binary star system. In addition to this main science goal, the
project allows us to explore photometric variability in century timescales and
possibly to detect some high ecliptic latitude asteroids.

\section{Method}

\begin{figure}
 \centering
 \includegraphics[width=\hsize]{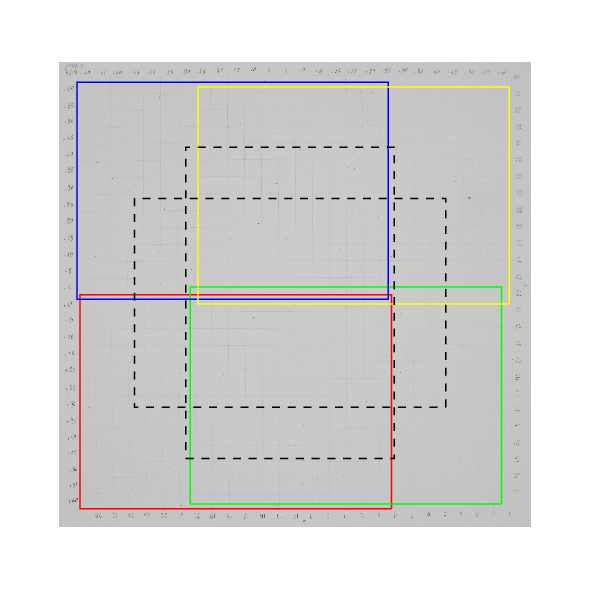}
 \caption{Schematic view of the two methods used to digitize each plate.  The
   four colored rectangles show the overlapping images that cover the whole
   plate. The two rectangles with dashed lines show the overlapping images that
   exclude the borders of the plate. The images taken with the former method are
   used in our final data analysis.}
 \label{fig:cducplate}
\end{figure}

\begin{table}
\caption{Details of the digitized plates. The last column states whether the
  plate is a single- (1) or triple-exposure (3) plate.}
\label{table:plates}      
\centering                          
\begin{tabular}{c c c c c}        
\hline\hline                 
Number & Observing date & R.A.   & Declination & Expo- \\    
       & [day/mn/year]  & [1900] & [1900]      & sure  \\   
\hline                        
  841 & 04/10/1896 & 20h00m & 40$\degr$00\arcmin & 1  \\
  844 & 06/11/1902 & 20h00m & 46$\degr$00\arcmin & 1  \\  
  883 & 03/10/1907 & 21h00m & 40$\degr$00\arcmin & 1  \\  
  886 & 18/08/1904 & 21h00m & 46$\degr$00\arcmin & 1 \\ 
  890 & 27/09/1904 & 21h10m & 40$\degr$00\arcmin & 1  \\ 
  892 & 05/11/1901 & 21h10m & 44$\degr$00\arcmin & 1  \\ 
  854 & 25/09/1903 & 20h15m & 45$\degr$00\arcmin & 3 \\
  887 & 13/10/1896 & 21h05m & 41$\degr$00\arcmin & 3 \\
  894 & 14/10/1896 & 21h15m & 41$\degr$00\arcmin & 3 \\
  896 & 19/08/1903 & 21h15m & 45$\degr$00\arcmin & 3 \\
\hline                                   
\end{tabular}
\end{table}

To digitize the CduC plates we used the digital camera Canon EOS~5Ds, which has
8736$\times$5856 image pixels, together with a Canon EF 100\,mm f/2.8L macro IS
USM lens. The pixel size is about 11\,$\mu$m corresponding to a nominal
resolution of about 0.68\arcsec\ per pixel.  Digital cameras for the consumer
market, such as the Canon EOS 5Ds, have a low-pass filter (so called anti-alias
filter) in front of the sensor. The filter prevents the Moire effect in an image
by effectively smoothing an image slightly.  In order to have the best possible
resolution the filter and the associated optics have been removed by JTW 
Astronomy\footnote{http://www.jtwastronomy.com} and replaced with an optically
polished clear glass filter.  This also required a small movement of the
detector array to accommodate to the change of the optical path.

Before taking the photos of the CduC plate, the glass is cleaned.  The camera is
attached to a Kaiser Reproduction Stand RS1/RA1 5510.  The glass plates are lit
from below by a LED illuminated light table Artograph LightPad A930.  To have
good enough resolution and to cover the whole plate, we take four partly
overlapping images of each plate so that the shorter side of the image sensor
covers half the height of a plate (1\degr) along Declination, and the longer
side is along Right Ascension (R.A.).  To check whether the most elongated stellar
profiles at the borders of the plates worsen the astrometric accuracy, we
digitize the plates also by taking two images that are centered at the center
of a plate and rotated 90\degr\ with respect to each other.  In this case, we
exclude the borders of the plates, covering $\sim$50\% of the plate area.  These
two methods are schematically shown in Fig.~\ref{fig:cducplate}.  In both cases,
the central parts of a plate are imaged twice. We disable the corrections for
peripheral illumination and chromatic aberration in the camera software.  The
optical axis of the lens is set perpendicular to the surface of a plate by
adjusting the yaw of the camera with the help of the electronic level of the
camera.  The pitch of the camera is fixed by the stand and cannot be adjusted.

In our project we utilize position and proper motion values of stars given in
the Tycho-2 \citep{Hog2000}, UCAC5 (USNO CCD Astrograph Catalog)
\citep{Zacharias2017}, HSOY (Hot Stuff for One Year) \citep{Altmann17}, PMA
\citep{Akhmetov2017} and {\it Gaia} TGAS (Tycho-{\it Gaia} Astrometric Solution)
\citep{Michalik2015} catalogs. The UCAC5, HSOY, and PMA are catalogs between the
first data release of {\it Gaia}, the {\it Gaia} DR1 \citep{Brown2016,
  Prusti2016, Lindegren2016}, and ground-based astrometry. We limit the UCAC5,
HSOY, and PMA catalogs to those stars that have {\it Gaia} G-band magnitude less
than 16\,mag.  In {\it Gaia} TGAS, the derivation of proper motion and parallax
values after only about one year of {\it Gaia} observations was made possible by
a joint solution of the Tycho-2 catalog with the {\it Gaia} DR1 data.  The
precisions of the proper motions in the Tycho-2, UCAC5, HSOY, and PMA catalogs
are 2.5\,mas/yr (all stars), 1-2\,mas/yr ($R=11-15$\,mag), $\la2$\,mas/yr ({\it
  Gaia} G-band magnitude $<15$\,mag), and 2-5\,mas/yr ({\it Gaia} G-band
magnitude 10-17\,mag), respectively.  For {\it Gaia} TGAS data the standard
uncertainties are $\sigma(\mu_{\alpha\star})=1.1$\,mas/yr and
$\sigma(\mu_\delta)=0.87$\,mas/yr (all primary sources).

The plates digitized by us are from a region of the sky where the density of the
{\it Gaia} TGAS stars is at its highest (R.A.$\approx$19h--22h). The typical total
number of stars on a CduC plate that can be identified with a star in a
reference star catalog is 350 stars with respect to the {\it Gaia} TGAS and
Tycho-2 catalogs, and 1713 stars with respect to the UCAC5, HSOY, and PMA
catalogs.  Table~\ref{table:plates} gives details of the digitized plates.

We use the SExtractor program \citep{Bertin1996} to search for the stars and to
derive their position and flux. SExtractor uses an iterative process to
determine the position of a source on an image, and the process works equally
well for circular or elliptical objects.  We use the SCAMP program
\citep{Bertin2006} to compute the astrometric solutions of the images.

\section{Data analysis}

\subsection{Removal of Bayer pattern}

Images of the CduC plates are saved in a 14-bit RAW format on a hard disk, then
converted into a TIFF format using
dcraw\footnote{http://www.cybercom.net/$\sim$dcoffin/dcraw.} program, and then
converted into a FITS format for data analysis. A scalar bias value is
subtracted from the images. Because the camera has a color sensor, each
2$\times$2 sub-array of a sensor has one pixel for red, one pixel for blue, and
two pixels for green light (Bayer color filter array).  We have investigated two
ways to remove the Bayer pattern. Firstly, we can use the built-in interpolation
options of the dcraw software. In this case, the missing pixels of each color
are interpolated, and we get complete images for each of the three colors. The
final image is then obtained by taking a weighted mean of the three
images. Secondly, we can read the RAW data without any built-in interpolation
within dcraw. Then, the mean values of the red and the blue pixels are scaled to
the mean value of the green pixels over the whole image.  These two methods give
practically the same results for the astrometry, but we have used the latter
method for our images. In this method, the scaling factors are determined mostly
by the background pixels, which form the vast majority of an image. This method
works as long as the spectrum of the light coming through a plate stays the
same.  Some plates have yellowish regions at their borders where scaling
residuals can be seen at the $\sim$5\% level. Furthermore, the plateaus at the
centers of some stars show scaling residuals at the $\sim$1\% level. The final
astrometric accuracy is not sensitive to the accuracy of the scaling correction;
changing the scaling factors for any color by $\pm$10\% increases the residuals
of the astrometry only by about 0.02\arcsec.

\subsection{Removal of grid lines}

When searching for stars in an image with SExtractor, the image is first heavily
smoothed to derive the background value at each pixel. Before smoothing, the
grid lines have to be removed. Our procedure is the following. Suppose we want
to remove the lines running along the y-axis. We take a mean value over each
column, obtaining a vector giving the shapes of the grid lines with a very good
signal-to-noise ratio as a function of the $x$-axis of an image.  Finding the
positions of the grid lines is then finding the positions where the derivative
of the vector changes from large positive values to large negative values. This
is a robust method because it does not depend on the absolute value of the
intensity. We then linearly interpolate the pixels which are $\pm$seven pixels
around the center of the grid line, and the interpolated pixels are flagged as
bad pixels. The flagged pixels constitute about 7\% of the total image area.
All the stars that have a bad pixel within their stellar profile are discarded.

\subsection{Astrometric solution}

The astrometric and photometric analysis of each plate consists of the following
steps:\newline i) Preliminary astrometry for the image is obtained with the
Astrometry.net software \citep{Lang2010}.  The purpose of this stage is to get a
preliminary astrometry that SCAMP software can successfully use as a starting
point in the refinement of the astrometry (step iv below).\newline ii) the
SExtractor software is used to detect the objects in the image and to derive
their fluxes and pixel positions.\newline iii) The coordinates of the stars in
the external catalogs are transformed to the epoch of the plate to be used as
reference stars for astrometry. The transformation is made as described in
Sect.~1.5.5 of Volume~1 of \citet{ESA1997}.  Also the uncertainties of the
transformed coordinates are computed.\newline iv) The SCAMP software is used to
compute astrometric projection parameters by using the pixel positions of stars
from step ii and the reference stars from step iii above, after dividing each
image into two sub-images along the shorter side of the image array.  SCAMP
derives a polynomial distortion model for an image by minimizing a weighted
quadratic sum of the differences in the coordinates (longitude and latitude)
between the stars in the plate and the reference stars.  A distortion of an
image can be visualized as a variation of the pixel scale as a function of
coordinates. The astrometric fit by SCAMP in the sub-images is computed by using
an average of 78 reference stars in the case of the {\it Gaia} TGAS and Tycho-2
catalogs, and an average of 181 stars in the case of the UCAC5, HSOY, and PMA
catalogs.  Due to the relatively low number of reference stars in our
sub-images, we use a low, second-order polynomial in the astrometric fit as a
starting point. Then, based on visual examination of check-plots produced by
SCAMP and the value of $\chi^2$ of the astrometric fit, we use a third-order
polynomial if necessary.  In the case of a second-order polynomial, the equation
for the longitudinal distortion has the form
\begin{equation}
x'=a+bx+cy+dx^2+exy+fy^2
,\end{equation}
where $x'$ is the distortion-corrected longitude, $x$ and $y$ are uncorrected
longitudinal and latitudinal offsets from the distortion-center origin, and $a$
to $f$ are constants. A similar equation applies for the latitude, by
interchanging $x$ and $y$ \citep{Shupe2005}.\newline v) The distortion
parameters computed by SCAMP are fed to the header of the image and SExtractor
is run again to derive the final celestial coordinates of the objects at the
epoch of each plate.

\subsection{Triple exposures}

We use SExtractor to detect the star images in the triple exposures. The
location of each star image relative to the two closest star images is used to
deduce which star images are members of a certain triplet. In this way, we
detect such triplets where the individual star images are detected as distinct
objects by SExtractor. The star images with a visual magnitude $\la8$\,mag
remain undetected because their triplets cannot be deblended into three distinct
objects. However, the number of such star images is small.
 
We fit the triplets with three overlapping elliptical Gaussians with a
saturation parameter \citep{Dick1993}. The width of all the three Gaussians is
expected to be equal. The fitted function for a triplet has the form
\begin{eqnarray}
I_{i,j} = & B + \sum_{k=1}^3 A_{k} \exp \{ -\frac{1}{2}[\frac{1}{1-t^2}
 ((\frac{x_{ij}-x_{ck}}{\sigma_x})^2\nonumber\\
 & +(\frac{y_{ij}-y_{ck}}{\sigma_y})^2
 -2t(\frac{x_{ij}-x_{ck}}{\sigma_x})(\frac{y_{ij}-y_{ck}}{\sigma_y}))]^s \}
,\end{eqnarray}
where $B$ is the value of the background, $x_{ij}$ and $y_{ij}$ are the pixel
coordinates, $x_{ck}$ and $y_{ck}$ are the center coordinates of the $k$th
Gaussian function, $A_{k}$ is the peak intensity of the Gaussian function,
$\sigma_x$, $\sigma_y$ and $t$ are widths and orientation of the elliptical
Gaussian functions, and $s$ is the flattening (saturation) parameter.  The
center position of the triplet in the pixel coordinates, $(x,y)$, is determined
as a mean value of the $x$- and $y$-positions of the three Gaussians. Then, we
follow the analysis steps given in Sect.~3.3 above, excluding step ii.

\subsection{Removal of instrumental distortions}\label{RemInsDis}

\begin{figure}
 \centering
 \includegraphics[width=\hsize]{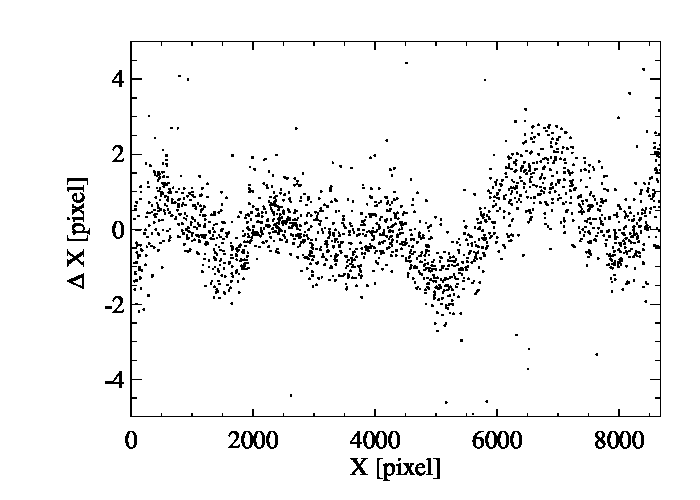}
 \caption{Difference, along the $x$-axis of an image, between the modeled
   and true coordinates of the artificial reference stars printed on a
   paper. The wavy relation is caused by uneven movement of a paper and/or a
   drum in a printer.}
 \label{fig:printer-deviat}
\end{figure}

The distortions caused by our instrumental setup include the distortions by the
camera plus lens combination, and the distortions caused by the
non-perpendicularity of the optical axis of the lens relative to the plate.  Our
procedure for removing these distortions is the following:\\ i) Make a fake,
random stellar field and print it on a paper. Also, make a corresponding star
catalog in a format that SCAMP can read.\\ ii) Take several images of the field
in a dither pattern, that is, move the paper slightly between the
exposures.\\ iii) Use SExtractor to derive the positions of the artificial
stars.\\ iv) Use SCAMP to derive the distortions for each image, based on the
positions derived in the previous step, and take a mean value of the distortion
parameters.\\ v) Use the SWarp program \citep{Bertin2002} to correct the CduC
images for the instrumental distortions.

Currently, our procedure for determining and correcting the distortions caused
by our instrumental setup (Sect.~\ref{RemInsDis}) is hampered by two facts:\\ i)
Consumer-grade laser printers suffer from an uneven movement of the drum
and/or paper at a level that is significant for us. This is illustrated in
Fig.~\ref{fig:printer-deviat}, which shows the one-dimensional difference, along
the $x$-axis of an image, between the modeled and true coordinates of the
artificial reference stars printed on a paper. The $x$-axis is along the
movement of a paper inside the printer. The wavy relation is not seen along the
$y$-axis, perpendicular to the movement of a paper.\\ ii) A proper determination
of the distortion requires an average over several distortion maps of an
artificial stellar field.  However, we have found no way to derive an average of the
distortion parameters when processing simultaneously several images of the
artificial stellar field using SCAMP.  Due to the above-mentioned facts, a
correction for the instrumental distortion is not implemented into our current
data analysis. With our method the instrumental distortion is simply removed as
part of astrometric distortion done separately for each plate.

\subsection{Photometric calibration}

The spectral response of our plates is not known, but the response peaks at blue
wavelengths. Therefore, we make photometric calibration for each plate
separately by forming a relation between the logarithm of the total flux of a
star on a CduC plate and its Tycho-2 $B_T$ magnitude.

\section{Results}

\begin{figure}
 \centering
 \includegraphics[width=\hsize]{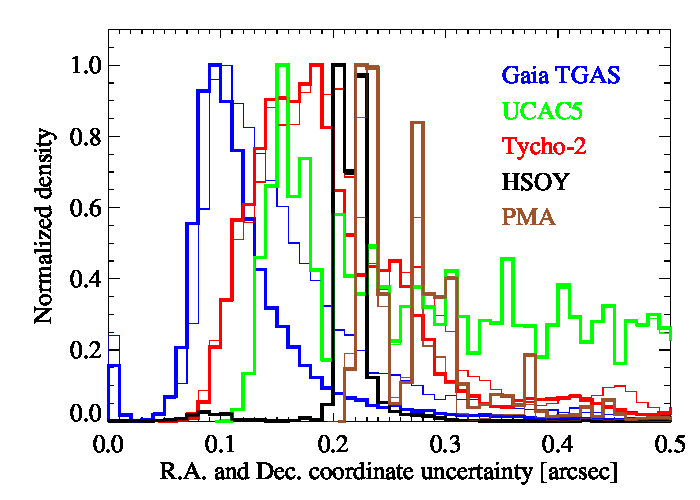}
 \caption{ Histograms of uncertainty of R.A.\ and Declination of
   the {\it Gaia} TGAS (blue), UCAC5 (green), Tycho-2 (red), HSOY (black), and
   PMA (brown) reference stars, after transforming their coordinates into the
   epochs of our plates. The thin line is for R.A., the thick line is for
   Declination. Only the stars within the areas of our plates are included.}
 \label{fig:sigmaCats}
\end{figure}

\begin{figure}
 \centering
 \includegraphics[width=\hsize]{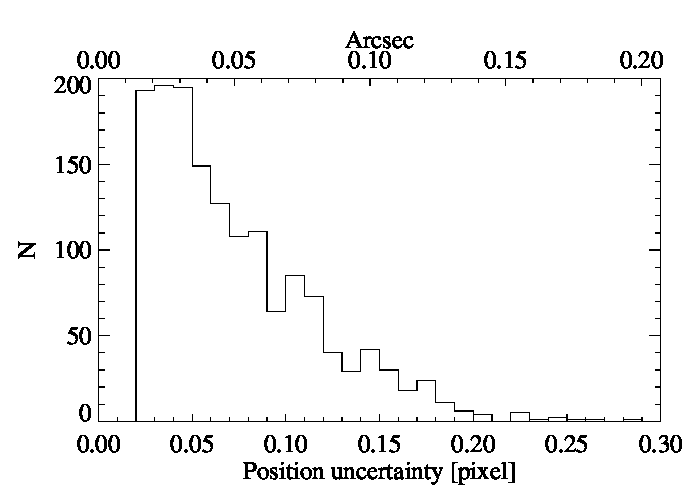}
 \caption{Example of a histogram of standard deviations of positions for the
   stars on one of our single-exposure image, as measured by SExtractor.  The
   histogram is a mean value of the deviations along the $x$- and $y$-axis of an
   image.}
 \label{fig:PosErrorHisto}
\end{figure}

\begin{figure}
 \centering
 \includegraphics[width=\hsize]{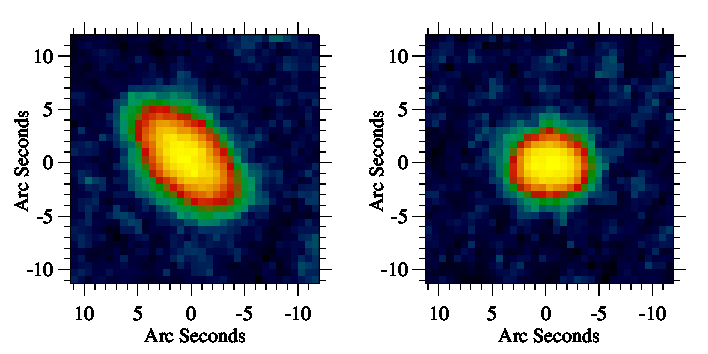}
 \caption{Examples of stellar profiles on our images. The left and right panels
   show profiles at a corner and center of a plate, respectively. The Tycho-2
   $B_T$ magnitudes of the stars are 12.8 (left panel) and 12.5 (right panel).}
 \label{fig:StellarProfiles}
\end{figure}

The histograms of uncertainty of R.A.\ ($\sigma(\alpha \cos(\delta))$)
and Declination ($\sigma(\delta)$) of the {\it Gaia} TGAS, UCAC5, Tycho-2, HSOY,
and PMA reference stars within the areas of our plates, after transforming their
coordinates into the epoch of each plate, are shown in Fig.~\ref{fig:sigmaCats}.
The error analysis includes the uncertainties of the coordinates (at the epoch
of each catalog), the uncertainties of the proper motion values, and, for {\it
  Gaia} TGAS data, also the correlations between the observational
parameters. Both for the {\it Gaia} TGAS and Tycho-2 data, the uncertainty is
greater along the R.A.\ due to the scanning strategy of the satellites.
To check whether the obtained astrometric accuracy is limited by the proper
motion uncertainties of the reference stars, we compute the astrometric
solutions also when using only those {\it Gaia} TGAS, UCAC5, HSOY, and PMA
reference stars which have uncertainties of the coordinates less than
0.15\arcsec, 0.20\arcsec, 0.27\arcsec,\ and 0.32\arcsec\ at the epoch of each
plate, respectively.

Figure~\ref{fig:PosErrorHisto} gives an example of the standard deviations of the pixel
positions of stars on an image, as computed by SExtractor.  The deviations are
mean values of the (very similar) deviations along the $x$- and $y$-axis of an
image.  Figure~\ref{fig:StellarProfiles} shows examples of the stellar profiles of a
star at the corner and center of a CduC plate, with Gaussian FWHM (Full Width at 
Half Maximum) sizes of
$\sim6\arcsec \times 10\arcsec$\ and $\sim7\arcsec$, respectively. Thus, within
the stellar profiles there may be dimmer, unseen stars, which move the
photo-center of the star seen on a plate away from its expected position.  This
effect is more serious at the corners of the plates because the stellar profiles
are larger there.

\subsection{Astrometric distortion}

\begin{figure}
 \centering
 \includegraphics[width=0.35\textwidth, angle=270]{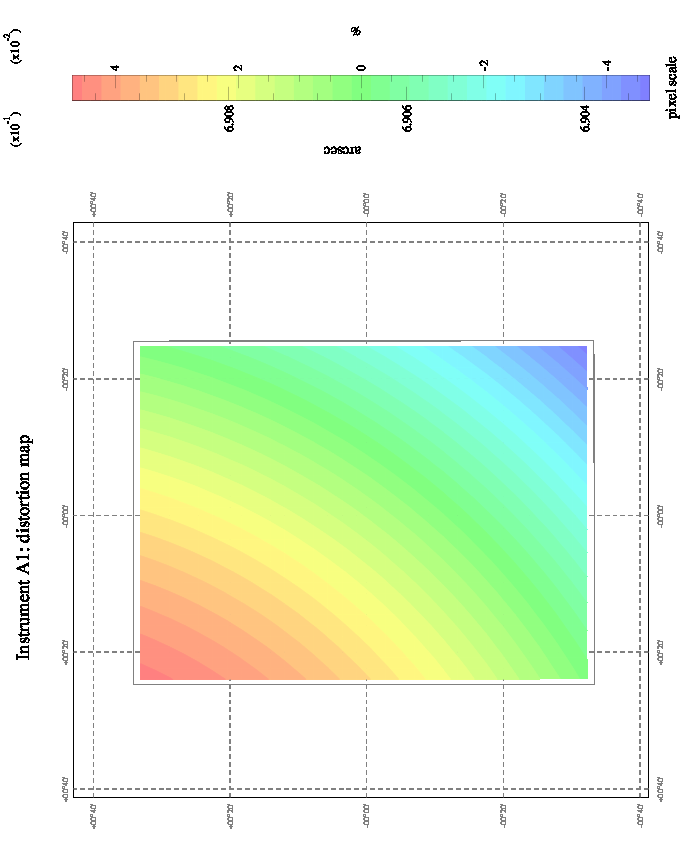}
 \caption{Example of a distortion map produced by an astrometric fit when
   using {\it Gaia} TGAS reference stars. The map is for one of the sub-images
   of the plate \#892.  The color bar gives the pixel scale of the image in
   absolute and percentual scales. The plot was produced by SCAMP.}
 \label{fig:distos}
\end{figure}

An example of a distortion map is shown in Fig.~\ref{fig:distos}. Values of
distortions over an image vary from about $\pm0.02\%$ to $\pm0.4\%$. We expect
that the distortion map of the refractor itself used to image the plates has the
largest pixel scale coinciding with the optical axis, that is,\ at the center of a
plate.  While some of our images do show such a distortion map, as the one shown
in Fig.~\ref{fig:distos}, we believe that in most cases the distortion maps are
dominated by distortions caused by the non-perpendicular orientation of the optical
axis of the camera relative to the plate.

For plates \#841 and \#890, a second-order polynomial cannot properly fit the
astrometric distortions. This is depicted, for example, by a skewed distribution of the
difference between the fitted and cataloged coordinates of the reference stars,
at the epoch of a plate. Using a third-order polynomial produces a non-skewed
distribution. By visually checking the distortion map produced by a third-order
polynomial fit we confirm that distortion changes over scales that are larger
than the typical distance between the reference stars, thus distortion is not
over-fitted.

\subsection{Astrometric analysis}

Our astrometric analysis includes the following steps; i) Computing astrometry
for quality checking of each plate, ii) Computing astrometry for each plate
that is of good quality and correcting the stellar coordinates for possible
biases, and iii) evaluating the astrometric fits. These steps are discussed in
the next three sections. Whenever we correlate the source coordinates on two
lists we use a search radius of 2\arcsec.

\subsubsection{Astrometric fit for quality checking}

\begin{figure*}
 \centering
 \includegraphics[width=\hsize]{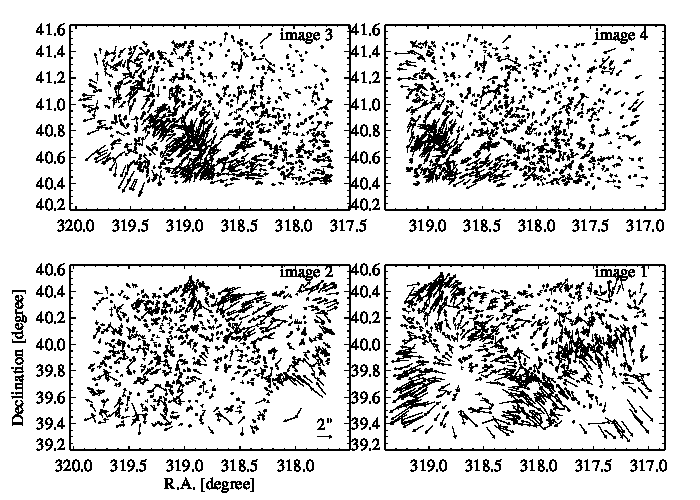}
 \caption{Large-scale astrometric vector residuals.  For each UCAC5 reference
   star which is used to fit the astrometry of an image, the arrow gives the
   magnitude and direction of the difference between the coordinates of the
   star, at the epoch of the plate, and the fitted coordinates of the
   corresponding star on the CduC image. The data is shown separately for each
   of the four partly overlapping images of plate \#890. The scale is shown
   in the lower right corner of image 2.}
 \label{fig:sigmaVectors}
\end{figure*}

The results in this section are given in the case of covering the plates with
four overlapping images (see Fig.~\ref{fig:cducplate}), and by simultaneously
processing all the partly overlapping eight sub-images of each plate with SCAMP.
The accuracy of the astrometric solutions for the plates are given in
Table~\ref{table:astrometry} as RMS (Root Mean Square) deviations of the 
astrometric fits along the
axes of the images (dAXIS1 and dAXIS2 in the notation by SCAMP, along Right
Ascension and Declination in our case, respectively). The internal calibration
represents the field-to-field calibration of the overlapping fields, while the
external calibration represents the calibration relative to the reference
stars. For the {\it Gaia} TGAS, UCAC5, HSOY, and PMA data, the results are given
both for the case of using all the stars, and using only those stars for which
the coordinate uncertainties, at the epoch of the plate, are less than the limit
given.  In the latter case, the accuracy of astrometry is marginally improved
for the {\it Gaia} TGAS and UCAC5 data.  However, the improvement in astrometry
is only about 0.02-0.03\arcsec, and the number of the reference stars is further
reduced.  Therefore all the further data analysis is done only for the case of
using all the reference stars in the external catalogs.

As discussed in Sect.~2 and shown in Fig.~\ref{fig:cducplate}, the plates have
been digitized also by taking two overlapping images, centered at the center of
the plate, and excluding the borders of the plate.  In that case the mean values
of dAXIS1 and dAXIS2 for the single-exposure plates, when using UCAC5 reference
stars, are 0.17\arcsec\ and 0.16\arcsec. These values are to be compared with
the corresponding values in Table~\ref{table:astrometry}, 0.21\arcsec\ and
0.20\arcsec. Thus, the accuracy of astrometry is improved by $\sim$0.04\arcsec when
excluding the borders of the plates.

Table~\ref{table:astrometry} shows that the deviation of the astrometric
solution differs between the plates by a factor of about two. Particularly, the
values of dAXIS and $\chi^2$ for plate \#890 are clearly larger than the
values for the other plates. Visual inspection of the images shows no difference
in the quality between these plates.  Furthermore, the uncertainties of the
pixel positions of stars are similar for all the plates.  To check whether the
astrometric residuals show structure on a large scale, we make an astrometric
fit for each full image (not dividing the image into two sub-images), using a
second-order polynomial in the fit, and plot the astrometric vector residuals.
Figure~\ref{fig:sigmaVectors} shows the astrometric residuals for each image of
 plate \#890.  There are regions where the direction and magnitude of the
residuals changes abruptly, such as in the north-west corner of image 2. The
vectors are not oriented in a way that is expected for a magnitude equation
(pointing radially outwards from the center of the plate), except at the
south-west corner of image 1.  We conclude that the intrinsic quality of 
plate \#890 is lower than that of the other plates, possibly due to a
large-scale deterioration of the emulsion. Plate \#841 shows a similar
behavior, but with smaller values of residuals.  Plate \#890 is included in
the results of Table~\ref{table:astrometry}, but further data analysis is done
without plate \#890.

\subsubsection{Biases in the stellar coordinates}

\begin{figure}
 \centering
 \includegraphics[width=\hsize]{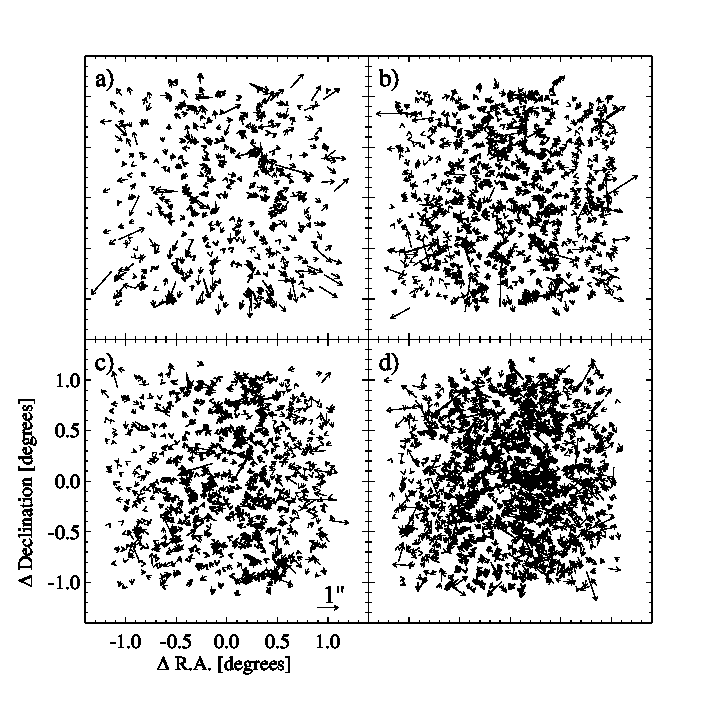}
 \caption{ Vector residuals for the UCAC5 reference stars in the single-exposure 
   plates, within four different magnitude ranges. The {\it Gaia}
   G-band magnitudes of the ranges are $<10$\,mag (panel a), 11.0-11.5\,mag
   (panel b), 12.0-12.25\,mag (panel c), and 13.0-13.25\,mag (panel d). }
 \label{fig:mag-equat-UCAC}
\end{figure}

\begin{figure}
 \centering
 \includegraphics[width=\hsize]{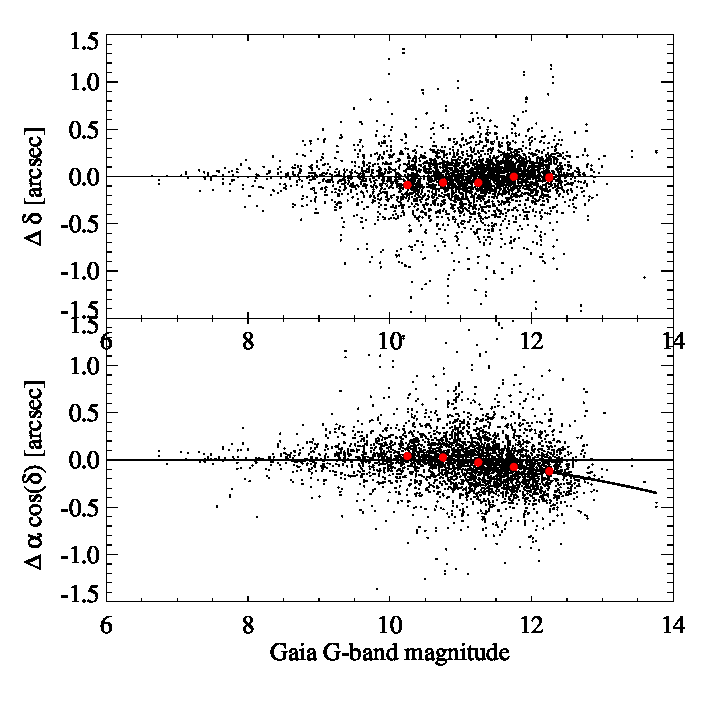}
 \caption{ Difference between the fitted and cataloged coordinates of all
   the {\it Gaia} TGAS stars in our single-exposure plates as a function of the
   {\it Gaia} G-band magnitude. Mean values over 0.5 magnitude bins are shown as
   red dots.  The lower panel is for R.A., the upper panel is for
   Declination. Overplotted in the lower panel is a second-order polynomial fit for
   magnitudes $>11$\,mag. }
 \label{fig:sigma-vs-magnitude}
\end{figure}

\begin{figure}
 \centering
 \includegraphics[width=\hsize]{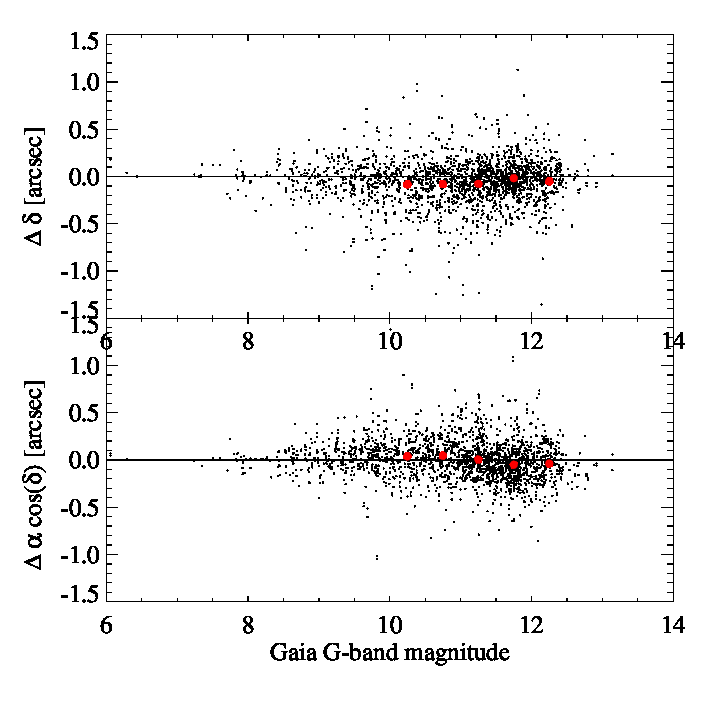}
 \caption{ Difference between the fitted and cataloged coordinates of all
   the {\it Gaia} TGAS stars in our triple-exposure plates as a function of the
   {\it Gaia} G-band magnitude. Mean values over 0.5 magnitude bins are shown as
   red dots.  The lower panel is for R.A., the upper panel is for Declination. }
 \label{fig:sigma-vs-magnitude-triple}
\end{figure}

\begin{figure}
 \centering
 \includegraphics[width=\hsize]{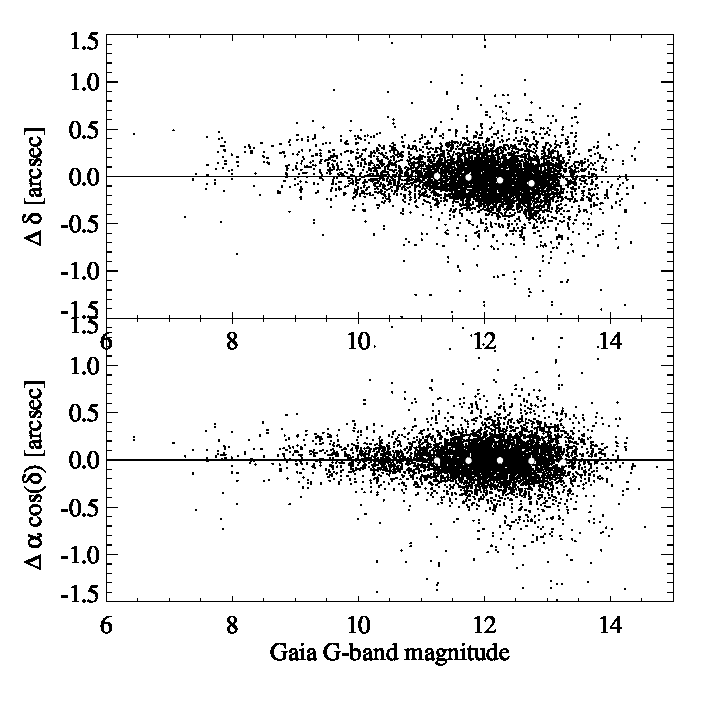}
 \caption{ Difference between the fitted and cataloged coordinates of all
   the Tycho-2 stars in our plates (both single- and triple-exposure) as a
   function of the {\it Gaia} G-band magnitude. Mean values over 0.5 magnitude
   bins are shown as white dots. The lower panel is for R.A., the upper panel is
   for Declination. }
 \label{fig:sigma-vs-magnitude-Tycho}
\end{figure}

\begin{figure}
 \centering
 \includegraphics[width=\hsize]{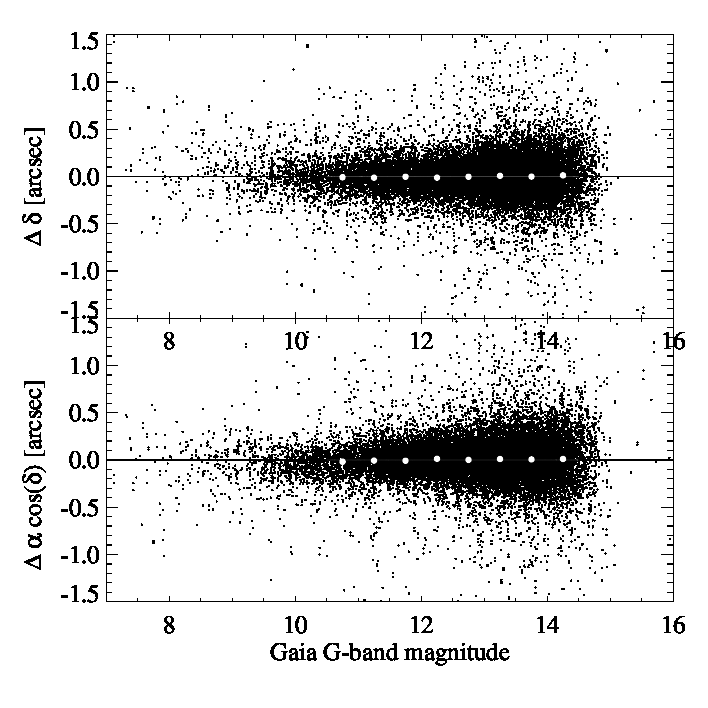}
 \caption{ Difference between the fitted and cataloged coordinates of all
   the UCAC5 stars in our single-exposure plates as a function of the {\it Gaia}
   G-band magnitude. Mean values over 0.5 magnitude bins are shown as white
   dots.  The lower panel is for R.A., the upper panel is for Declination. }
 \label{fig:sigma-vs-magnitude-UCAC5}
\end{figure}

\begin{figure}
 \centering
 \includegraphics[width=\hsize]{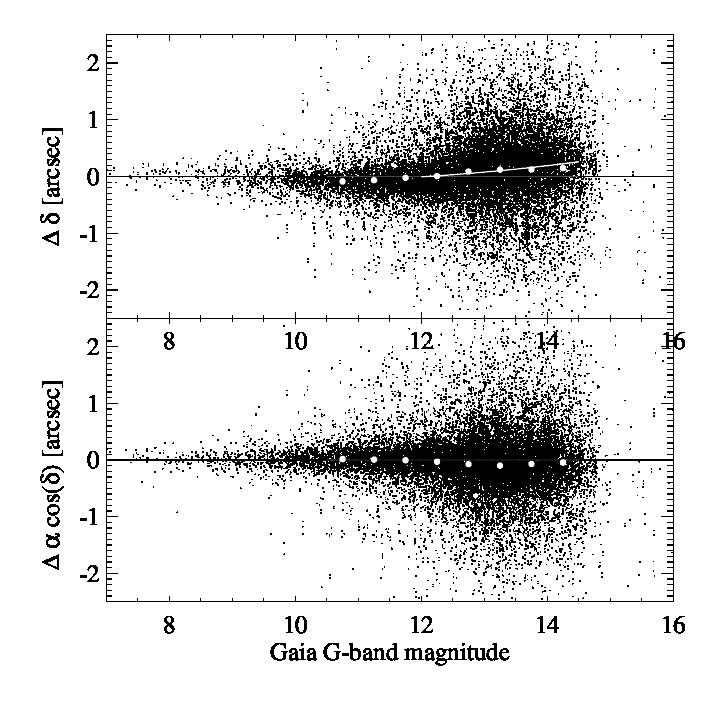}
 \caption{ Difference between the fitted and cataloged coordinates of all
   the HSOY stars in our single-exposure plates as a function of the {\it Gaia}
   G-band magnitude. Mean values over 0.5 magnitude bins are shown as white
   dots.  The lower panel is for R.A., the upper panel is for
   Declination. Overplotted in the upper panel is a second-order polynomial fit for
   magnitudes $>12$\,mag. }
 \label{fig:sigma-vs-magnitude-HSOY}
\end{figure}

\begin{figure}
 \centering
 \includegraphics[width=\hsize]{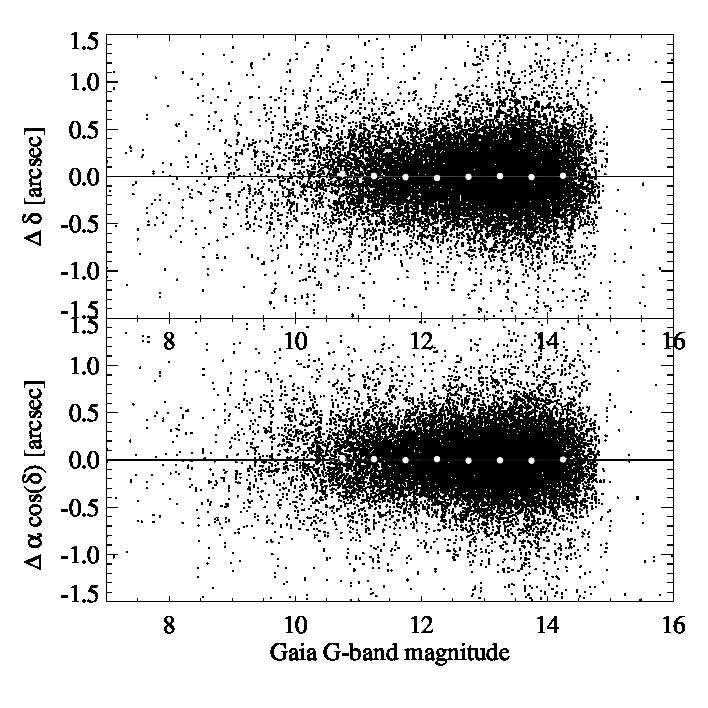}
 \caption{ Difference between the fitted and cataloged coordinates of all
   the PMA stars in our single-exposure plates as a function of the {\it Gaia}
   G-band magnitude. Mean values over 0.5 magnitude bins are shown as white
   dots.  The lower panel is for R.A., the upper panel is for Declination. }
 \label{fig:sigma-vs-magnitude-PMA}
\end{figure}

\begin{figure}
 \centering
 \includegraphics[width=\hsize]{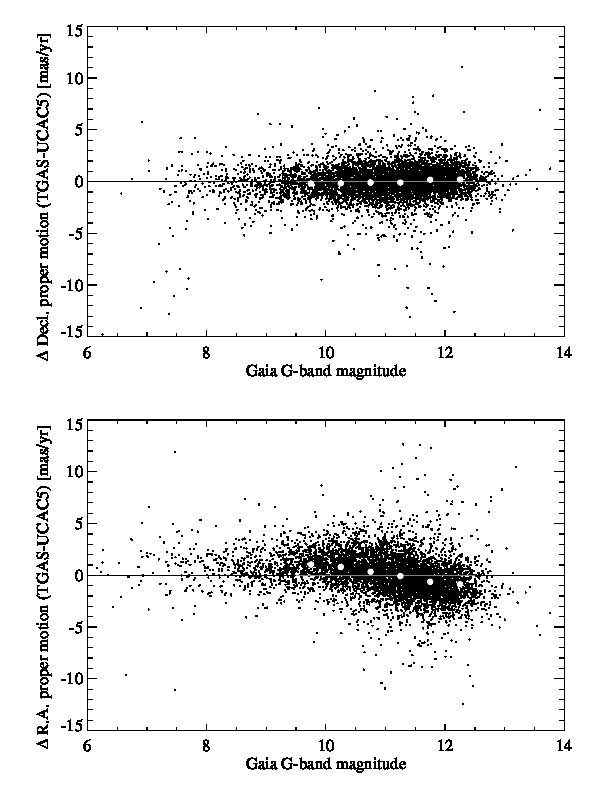}
 \caption{ Differences of the proper motions given in the {\it Gaia} TGAS
   and UCAC5 catalogs as a function of magnitude for the stars within our
   plates.  Mean values over 0.5 magnitude bins are shown as white dots.  The
   lower panel is for R.A., the upper panel is for Declination. }
 \label{fig:Gaia-UCAC-prop-mot}
\end{figure}

The results in this section are given in the case of covering the plates with
four overlapping images (see Fig.~\ref{fig:cducplate}), and after individually
processing all the partly overlapping eight sub-images of each plate with SCAMP.
It is known that the photographic plate material may suffer from a
magnitude-dependent systematic bias in the positions of the stars, the so-called
magnitude equation (e.g.,\ \citet{Girard1998, Vicente2010}).  It is caused by the
combination of asymmetric stellar profiles and the nonlinear response of
photographic emulsions. As a result, the photographic density profiles of stars
are skewed, so that the profiles of stars are more skewed for the brighter
stars, which are more affected by the nonlinearity of the emulsion. The
asymmetry itself can be caused by, for example,\ optical aberrations or tracking errors
during an exposure. We believe that for the CduC plates the asymmetry is
dominated by field curvature, which produces elongated stellar profiles with the
long axis of the profile pointing towards the center of the plate
\citep{OrtizGil1998} (see Fig.~\ref{fig:StellarProfiles}).

The differences between the fitted and cataloged coordinates (vector residuals)
of the UCAC5 stars in the single-exposure plates, within different magnitude
ranges, are shown in Fig.~\ref{fig:mag-equat-UCAC}. The magnitude of the
residuals is largest for the brightest stars, as expected for magnitude
equation. The residuals are pointed outward from the center of the plate,
similarly to the magnitude equation found by \citet{Vicente2010} for CduC
plates.  However, the number of stars is too low to determine an accurate
correction for magnitude equation in our plates.

Magnitude equation is barely visible with the {\it Gaia} TGAS reference stars.
It is possible that due to the lower number of {\it Gaia} TGAS reference stars,
the SCAMP is able to fit some of the distortion caused by magnitude equation.
Related to the magnitude equation, the distances between all the stars detected
in the single-exposure plates and their counterparts in the {\it Gaia} TGAS
catalog (i.e.,\ astrometric residuals) as a function of the {\it Gaia} G-band
magnitude are plotted in Fig.~\ref{fig:sigma-vs-magnitude}. At magnitudes
$\ga11$ the fitted R.A.\ coordinates of stars start to deviate from their
cataloged coordinates.  Overplotted in the lower panel of
Fig.~\ref{fig:sigma-vs-magnitude} is a second-order polynomial fit, which is
subtracted from the fitted R.A.\ coordinates of those stars that have {\it
  Gaia} G-band magnitude greater than 11\,mag.  The corresponding image for the
triple-exposure plates is in Fig.~\ref{fig:sigma-vs-magnitude-triple}. There is
a similar bias in the R.A.\ coordinates than for the single-exposure plates, but
the bias is not clear enough to be fitted and subtracted.

The astrometric residuals for the Tycho-2 stars as a function of the {\it Gaia}
G-band magnitude are plotted in Fig.~\ref{fig:sigma-vs-magnitude-Tycho}.  There
is a small bias in the residuals along Declination, with a maximum value of
about -0.07\arcsec. We do not correct this bias.
The astrometric residuals for the UCAC5 stars in the single-exposure plates as a
function of the {\it Gaia} G-band magnitude are plotted in
Fig.~\ref{fig:sigma-vs-magnitude-UCAC5}. There is no bias in the residuals. The
same is true for the triple-exposure plates.
The astrometric residuals for the HSOY stars in the single-exposure plates as a
function of the {\it Gaia} G-band magnitude are plotted in
Fig.~\ref{fig:sigma-vs-magnitude-HSOY}. The residuals along Declination are
biased, and overplotted in the upper panel is a second-order polynomial fit, which
is subtracted from the fitted Declination coordinates of those stars that have
{\it Gaia} G-band magnitude greater than 12\,mag. The residuals along R.A.\ show
a small bias but we do not correct it.  There is no clear bias in a similar plot
for the triple-exposure plates.
The astrometric residuals for the PMA stars in the single-exposure plates as a
function of the {\it Gaia} G-band magnitude are plotted in
Fig.~\ref{fig:sigma-vs-magnitude-PMA}. There is no bias in the residuals. The
same is true for the triple-exposure plates.

We conclude that the coordinates of the {\it Gaia} TGAS and HSOY reference
stars, at the epoch of the plates, are biased along R.A.\ and Declination,
respectively, as a function of magnitude.

We also check whether the difference between the fitted and cataloged
coordinates of the {\it Gaia} TGAS stars in our plates is correlated with the
properties of those stars (other than the correlation in
Fig.~\ref{fig:sigma-vs-magnitude}).  The properties include ellipticity of a
star, position angle of an elliptic stellar profile, width of a stellar profile,
pixel position of a star on a plate, and color of a star (Tycho-2 $B_T$ and
$V_T$ magnitudes). We find no correlation. The differences between the
coordinates of the Tycho-2 and {\it Gaia} TGAS stars, at the epoch of the
plates, do not show any correlation with the properties of the corresponding stars on a
CduC plate.

Figure~\ref{fig:Gaia-UCAC-prop-mot} shows the differences between the {\it Gaia}
TGAS and UCAC5 proper motions for the stars within our plates, as a function of
{\it Gaia} G-band magnitude. There is a bias in R.A.\ proper motions, with a
value of $\sim$1mas/yr for a {\it Gaia} G-band magnitude of $\sim$12. Within a
time span of 100\,years, this corresponds to a R.A.\ coordinate bias of
$\sim$0.1\arcsec, close to the value that we observe with {\it Gaia} TGAS
reference stars in Fig.~\ref{fig:sigma-vs-magnitude}.  We conclude that the bias
in the astrometric residuals of the {\it Gaia} TGAS stars along R.A.\ is
probably caused by a bias in the values of the R.A.\ proper motions in the {\it
  Gaia} TGAS catalog. An equivalent plot of the proper motion differences
between the HSOY and UCAC5 stars as a function of {\it Gaia} G-band magnitude
shows a bias along Declination, which could explain the bias seen in
Fig.~\ref{fig:sigma-vs-magnitude-HSOY}.

\subsubsection{Astrometric accuracy}

\begin{figure*}
 \centering
 \includegraphics[width=\hsize]{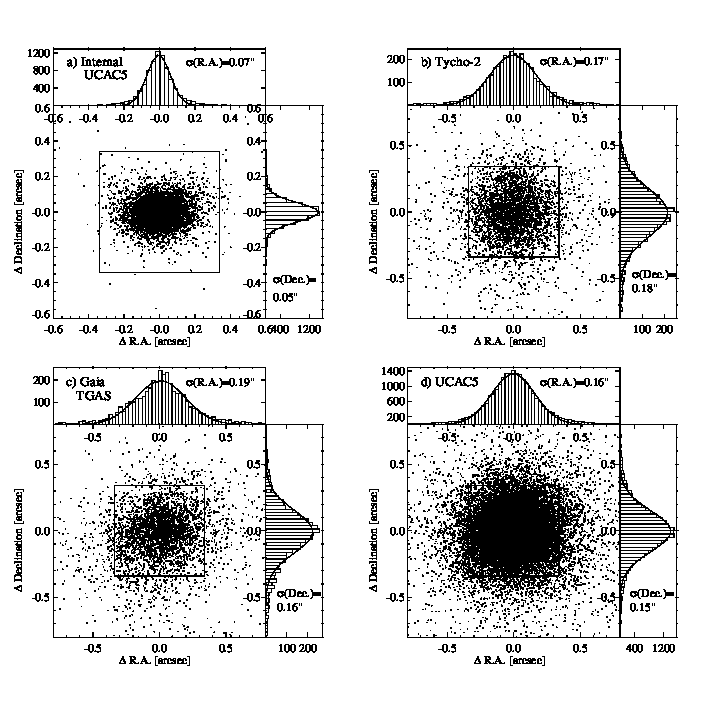}
 \caption{Astrometric results for the single-exposure plates.  a) The
   differences between the coordinates of the UCAC5 stars in the overlapping
   images at the epoch of each plate. b)--d) The differences between the
   coordinates of all the stars in the plates and the coordinates of the
   corresponding stars in the Tycho-2, {\it Gaia} TGAS, and UCAC5 catalog,
   respectively, at the epoch of each plate.  In all the panels, the size of an
   image pixel is shown as a box, Gaussian fits are plotted over the histograms,
   and the Gaussian standard deviations are given.}
 \label{fig:AstrometrySinglePlates}
\end{figure*}

\begin{figure*}
 \centering
 \includegraphics[width=\hsize]{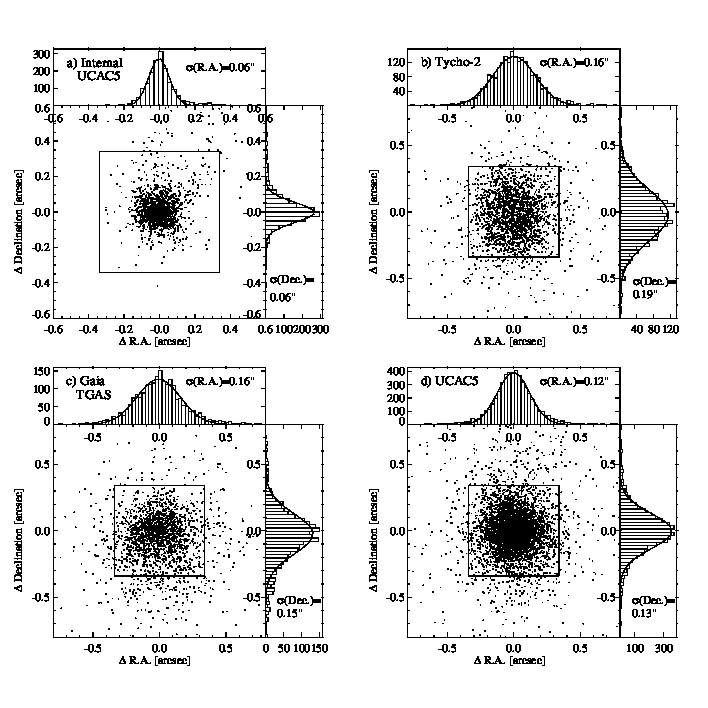}
 \caption{ Astrometric results for the triple-exposure plates.  a) The
   differences between the coordinates of the UCAC5 stars in the overlapping
   images at the epoch of each plate. b)--d) The differences between the
   coordinates of all the stars in the plates and the coordinates of the
   corresponding stars in the Tycho-2, {\it Gaia} TGAS, and UCAC5 catalog,
   respectively, at the epoch of each plate.  In all the panels, the size of an
   image pixel is shown as a box, Gaussian fits are plotted over the histograms,
   and the Gaussian standard deviations are given. }
 \label{fig:AstrometryTriplePlates}
\end{figure*}

\begin{figure}
 \centering
 \includegraphics[width=\hsize]{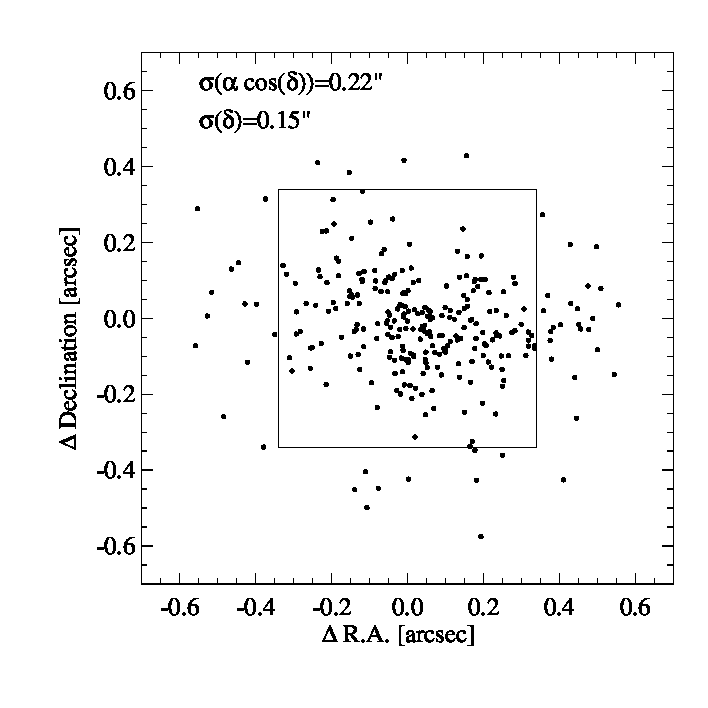}
 \caption{ Difference of the coordinates of the common stars in the partly
   overlapping plates \#892 and \#896. The size of an image pixel is shown as a
   box, and standard deviations are given. }
 \label{fig:OverlapPlates}
\end{figure}

\setcounter{table}{2}

\begin{sidewaystable*}
\caption{Astrometric accuracy of the CduC plates.  The deviations given are
  Gaussian 1-$\sigma$ deviations along R.A.\ and Declination (see
  Figs.~\ref{fig:AstrometrySinglePlates} and
\ref{fig:AstrometryTriplePlates}). The meaning of the colums: 
  1: plate type (single or triple exposure), 
  2-3: deviation of R.A.\ and Declination for UCAC5 reference stars in the overlapping images, 
  4-5: deviation of R.A.\ and Declination for Tycho-2 reference stars in all the images, 
  6-7: deviation of R.A.\ and Declination for {\it Gaia} TGAS reference stars in all the images, 
  8-9: deviation of R.A.\ and Declination for UCAC5 reference stars in all the images, 
10-11: deviation of R.A.\ and Declination for HSOY reference stars in all the images, 
12-13: deviation of R.A.\ and Declination for PMA reference stars in all the images.}
\label{table:astaccuracy}     
\centering    
\begin{tabular}{c c c c c c c c c c c c c}       
\hline\hline                 
Plate & $\sigma(\alpha_{\cos(\delta)})$ & $\sigma(\delta)$ & 
        $\sigma(\alpha_{\cos(\delta)})$ & $\sigma(\delta)$ & 
        $\sigma(\alpha_{\cos(\delta)})$ & $\sigma(\delta)$ & 
        $\sigma(\alpha_{\cos(\delta)})$ & $\sigma(\delta)$ & 
        $\sigma(\alpha_{\cos(\delta)})$ & $\sigma(\delta)$ &
        $\sigma(\alpha_{\cos(\delta)})$ & $\sigma(\delta)$ \\
type & overlap.  & overlap.      & Tycho-2 & Tycho-2 & {\it Gaia} & {\it Gaia} & UCAC5 & UCAC5 & HSOY & HSOY & PMA & PMA \\
     &    image    &    image    &         &         & TGAS & TGAS &       &       &      &      &     &     \\
\hline        
1 & 2 & 3 & 4 & 5 & 6 & 7 & 8 & 9 & 10 & 11 \\
\hline
Single & 0.07\arcsec & 0.05\arcsec & 0.17\arcsec & 0.18\arcsec & 0.19\arcsec & 0.16\arcsec & 0.16\arcsec 
       & 0.15\arcsec & 0.30\arcsec & 0.33\arcsec & 0.27\arcsec & 0.31\arcsec \\
Triple & 0.06\arcsec & 0.06\arcsec & 0.16\arcsec & 0.19\arcsec & 0.16\arcsec & 0.15\arcsec & 0.12\arcsec 
       & 0.13\arcsec & 0.17\arcsec & 0.20\arcsec & 0.26\arcsec & 0.29\arcsec \\
\end{tabular}
\end{sidewaystable*}

The astrometric accuracy of the final stellar coordinates can be evaluated in
three ways; i) the deviation between the coordinates of a star in two
overlapping images of the same plate, ii) the deviation of the coordinates
between all the stars detected and their counterpart in the reference star
catalog, at the epoch of each plate, and iii) the deviation of the stellar
coordinates in the overlapping plates. \\ i) The differences between the fitted
coordinates of the UCAC5 stars in the overlapping images of the single- and
triple-exposure plates are plotted in Figs.~\ref{fig:AstrometrySinglePlates}a 
and~\ref{fig:AstrometryTriplePlates}a, respectively.  The Gaussian standard
deviations of the differences are given in
Table~\ref{table:astaccuracy}.\\ ii) The differences between the fitted
coordinates of all the stars detected and their counterparts in the Tycho-2,
{\it Gaia} TGAS, and UCAC5 catalogs are plotted in
Figs.~\ref{fig:AstrometrySinglePlates}\,b-d 
and~\ref{fig:AstrometryTriplePlates}\,b-d, for the single- and triple-exposure
plates, respectively. The Gaussian standard deviations of the differences are
given in the Table~\ref{table:astaccuracy}.\\ iii) The differences of the fitted
coordinates of the stars which are common to the partly overlapping plates \#892
and \#896 are shown in Fig.~\ref{fig:OverlapPlates}.

The pixel positions of stars measured by SExtractor are Gaussian-weighted
centroids of stellar profiles. To check whether the accuracy of astrometry
depends on the method to determine the pixel positions of the stars, we perform
our analysis for some single-exposure plates also by fitting the stellar
profiles with a two-dimensional (2D) Gaussian function. The astrometric accuracy is then
practically the same as in the case of using SExtractor.

\subsection{Photometry}

\begin{figure}
 \centering
 \includegraphics[width=\hsize]{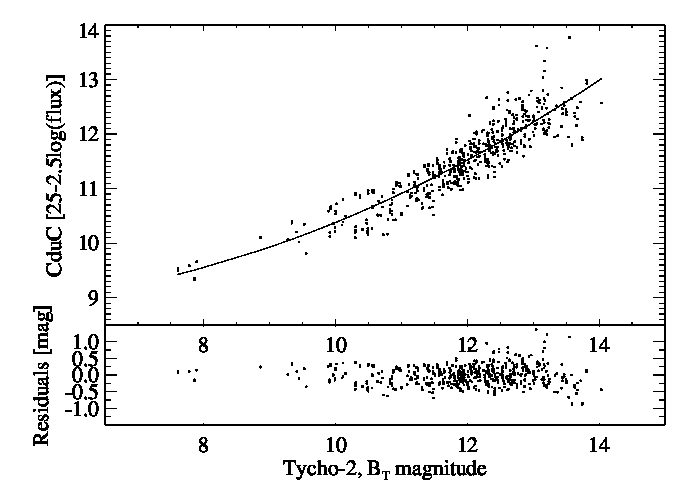}
 \caption{ Example of magnitude calibration. The upper panel shows the logarithm of
   the total flux of a star on a CduC plate as a function of the Tycho-2 $B_T$
   magnitude. The lower panel shows the residuals after the overplotted second-order
   fit has been subtracted from the CduC magnitudes. This example is for
   plate \#841.}
 \label{fig:photometry}
\end{figure}

\begin{figure}
 \centering
 \includegraphics[width=\hsize]{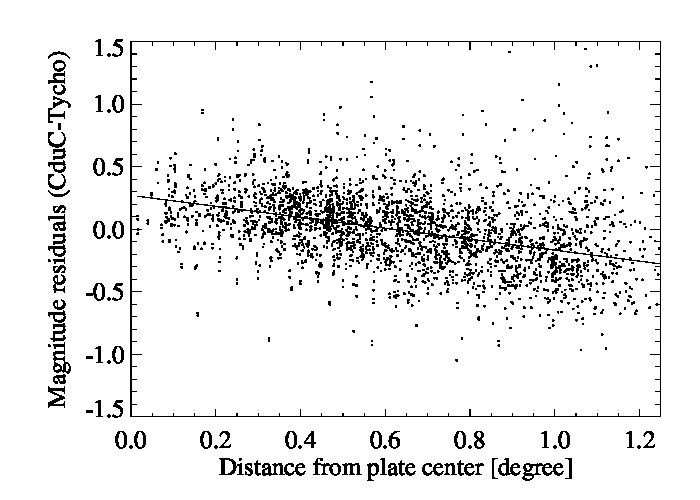}
 \caption{Magnitude residuals (CduC magnitudes minus Tycho-2 $B_T$
   magnitudes) as a function of the distance of the star from the center of the
   plate, for the single-exposure plates. A fit of a straight line to the data
   is shown.}
 \label{fig:magresiduals}
\end{figure}

\begin{figure}
 \centering
 \includegraphics[width=\hsize]{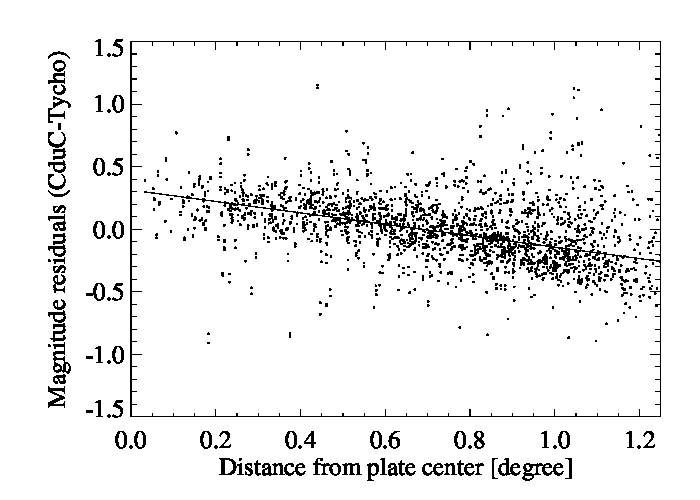}
 \caption{Magnitude residuals (CduC magnitudes minus Tycho-2 $B_T$
   magnitudes) as a function of the distance of the star from the center of the
   plate, for the triple-exposure plates. A fit of a straight line to the data
   is shown.}
 \label{fig:magresiduals-3}
\end{figure}

Figure~\ref{fig:photometry} shows an example of a relation between the total
flux of stars, measured from a CduC plate, and the Tycho-2 $B_T$ magnitudes. A
second-order polynomial is fitted to the relation and residuals are formed by
subtracting the fit from the CduC based magnitudes. The corresponding residuals
for all the stars in the single-exposure plates are shown in
Fig.~\ref{fig:magresiduals} as a function of the distance of the star from the
center of the plate. Near the center of the plate the stars are dimmer, while
near the edges of the plate the stars are brighter than expected. The trend in
Fig.~\ref{fig:magresiduals} is fitted with a straight line, the line is
subtracted from the residuals, and the standard deviation of the residuals,
$\sim$0.28\,mag, is used as a measure of photometric accuracy for the CduC
stars.  There is no correlation between the magnitude residuals and the colors
(Tycho-2 $B_T$ minus $V_T$ magnitudes) or the Tycho-2 $B_T$ magnitude of the
stars.
The corresponding plot for the triple-exposure plates is shown in
Fig.~\ref{fig:magresiduals-3}.  The photometric accuracy is $\sim$0.24\,mag for
the triple-exposure plates.

\subsection{Triple-exposure images}

\begin{figure}
 \centering
 \includegraphics[width=\hsize]{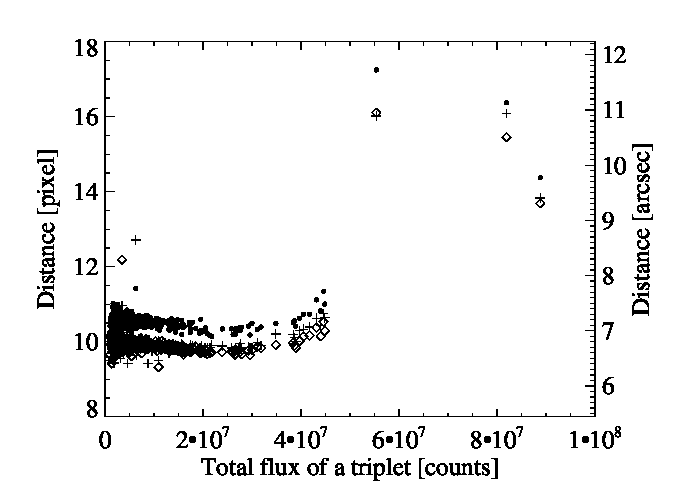}
 \caption{Adjacency effect on the triple-exposure plate \#854. For each star
   image of a triplet, forming a triangle, the distance of the star from the
   center of mass of the triangle is plotted as a function of the total flux of
   the triplet. The stars at the different corners of the triangle are plotted
   with a different symbol (plus sign, square, circle).}
 \label{fig:kostinsky}
\end{figure}

The results of the astrometric fit for the triple-exposure plates are in
Tables~\ref{table:astrometry} and~\ref{table:astaccuracy}. The accuracy of
astrometry is better than that of the single-exposure plates. This is expected
because the stellar positions are calculated as a mean value of three images of
a star.  This also means that the astrometric accuracy of single-exposure plates
is limited by the accuracy of the measured pixel positions of stars on a plate.

Figure~\ref{fig:kostinsky} shows for each of the three stars within a triplet
the distance of the star from the center of the triplet as a function of the
total flux of the triplet.  The plot is for plate \#854, including all the
stars detected by our automatic procedure plus three manually selected, highly
blended triplets not detected by our procedure (the three brightest stars).  The
star images at one of the corners of the asterisms (circles in
Fig.~\ref{fig:kostinsky}) are located further away from the center than the star
images at the other two corners. The reason for this is that stars in triplets
are not at the same distance from the center of the mass of a triplet.  The
Kostinsky effect (increase of distance of star images in the triplet relative to
each other as a function of stellar flux) is clearly detected.  The functional
form of the Kostinsky effect is similar to that found by \citet{Dick1993}
and \citet{OrtizGil1998}; above a certain flux threshold, the relative distances
of the stars in a triplet increase rapidly. In addition, our data show that for
low fluxes the distances slightly decrease as a function of flux.
We do not try to correct for this effect because the $x$/$y$ pixel coordinates
of each triplet, the center of mass of a triplet, is calculated as a mean value
of the $x$/$y$ coordinates, which is expected to stay stationary even if the
star images in the triplet move relative to the center of mass.

\section{Comparison with other CduC catalogs}

\citet{Rapaport2006} have analyzed 512 triple-exposure CduC plates, scanned with
the Cambridge APM (Automatic Plate Measuring) machine. They give a value of
$\sigma(\alpha\cos(\delta)) \simeq \sigma(\delta) \simeq 0.16\arcsec$ for the
precision of a coordinate of a single image of a star near the center of a
plate. Near the edges of a plate the precisions are
$\sigma(\alpha\cos(\delta))=0.17\arcsec$ and $\sigma(\delta)=0.20\arcsec$. When
considering the star triplets, the position uncertainty of the star of the
triplet is lower by $\sqrt{3}$, and consequently they give an anticipated value
of $\sim$0.11\arcsec\ for the astrometric standard errors.

To compare the positional accuracy obtained by \citet{Vicente2007, Vicente2010}
with our results, we quote their values for 'reference star residuals of
CduC catalog based on a comparison with the Tycho-2 positions at the epoch of
the plates', with standard deviations of
$\sigma(\alpha\cos(\delta))=0.22$\arcsec\ and
$\sigma(\delta)=0.24$\arcsec\ (Fig.~8 of \citet{Vicente2010}), including both
single- and triple-exposure plates.

A single triple-exposure plate was scanned with a microdensitometer by
\citet{OrtizGil1998}. They obtained astrometric accuracy of $\sigma(\Delta
x)=0.16$\arcsec\ and $\sigma(\Delta y)=0.13$\arcsec\ when fitting the whole
plate with ten reference stars.

The residuals in our survey, $\sigma(\alpha\cos(\delta))=0.17\arcsec$ and
$\sigma(\delta)=0.18\arcsec$ for the single-exposure plates in the case of
Tycho-2 reference stars, are about equal to or lower than the residuals in the
above-mentioned surveys. Our residuals for the triple-exposure plates,
$\sigma(\alpha\cos(\delta))=0.16$\arcsec\ and $\sigma(\delta)=0.19$\arcsec\ for
the Tycho-2 reference stars, are somewhat larger than those of
\citet{Rapaport2006}, but similar to those of \citet{OrtizGil1998}.  It is
important to note that the analyzed CduC plates were obtained with different
physical telescopes at different observatories. Therefore, astrometric
precisions not only stem from the analysis work.

Photographic magnitudes of stars on CduC plates have been obtained by
\citet{OrtizGil1998} with an accuracy of 0.09\,mag, by \citet{Lamareille2003}
with an accuracy of $\sim$0.2-0.4\,mag, and by \citet{Rapaport2006} with an
accuracy of 0.6\,mag. We obtain similar photometric accuracy, 0.28\,mag and
0.24\,mag for single- and triple-exposure plates, respectively.

\section{Discussion}

\begin{figure}
 \centering
 \includegraphics[width=\hsize]{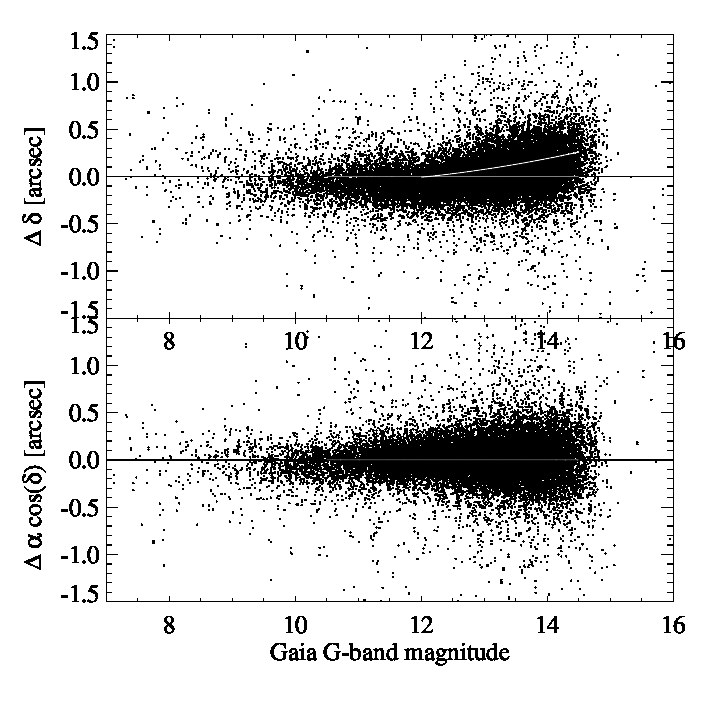}
 \caption{ Difference between the fitted and cataloged coordinates of all
   the UCAC5 stars in our single-exposure plates as a function of the {\it Gaia}
   G-band magnitude after we have modified the values of the Declination proper
   motions given in the UCAC5 catalog. The amount of the additive modification
   is given as a curved white line in units of $10$\,mas/yr.  The lower panel is
   for R.A., the upper panel is for Declination. }
 \label{fig:sigma-vs-magnitude-UCAC5-mod-pm}
\end{figure}

Each plate has been digitized in two ways; i) two overlapping images excluding
the borders of the plates, and ii) four overlapping images covering the whole
plate. The former method would be adequate to measure all the stars because the
CduC plates are overlapping so that no star is lost. Furthermore, the
former method gives astrometric residuals that are lower by $\sim$0.04\arcsec.
However, we prefer the latter method for the following reasons: i) The
difference in astrometry is most probably caused by magnitude equation, which
produces larger residuals at the borders of the plates (see
Fig.~\ref{fig:mag-equat-UCAC}). We can correct for this effect after we have
digitized more plates. And ii) If we exclude the borders of the plates, we lose
photometric information on the stars in the excluded regions.

The uncertainty of the coordinates of the {\it Gaia} TGAS stars, at the epoch of
the plates, is about two times lower than the uncertainty of the coordinates of
the Tycho-2 stars (Fig.~\ref{fig:sigmaCats}).  However, we find that the
accuracy of astrometry is similar for {\it Gaia} TGAS and Tycho-2 reference
stars, even after correcting for the bias in {\it Gaia} TGAS data
(Fig.~\ref{fig:sigma-vs-magnitude}).  We give three explanations for the
similarity:\\ i) It is explained by underestimated {\it Gaia} TGAS proper motion
uncertainties and/or overestimated Tycho-2 proper motion uncertainties, or by an
unknown distortion in the {\it Gaia} TGAS data.\\ ii) The pixel positions of
stars on the CduC plates, measured by us, are biased in the same way as the
coordinates of the Tycho-2 reference stars, at the epoch of the plates. This
biasing is probably caused by the large stellar profiles on photographic
plates. Within a stellar profile, there may be dim stars that move the
photocenter of the star on a plate away from the true position of the
star. Thus, the stellar positions measured manually about 100 years ago, and
digitally by us, are biased in the same way.  On the other hand, the proper
motions given in the Tycho-2 catalog largely depend on the stellar positions
from the Astrographic catalog.  Thus, the stellar positions in the Helsinki
section of the Astrographic catalog survey were used to derive the proper
motions given in the Tycho-2 catalog.  As a result, the positions of the stars
calculated by using the non-biased proper motions given in the {\it Gaia} TGAS
catalog do not give as good astrometry as would be expected based solely on the
uncertainties of the proper motion values.  On the other hand, we obtain the
best astrometry with the UCAC5 reference star catalog, which is based on higher
resolution ground-based CCD observations (and {\it Gaia} DR1 data). This suggests
that the large stellar profiles on photographic plates are not the reason for the
similar astrometric accuracy between {\it Gaia} TGAS and Tycho-2 data.\\ iii)
The intrinsic accuracy of the plates is worse or about the same as the accuracy
of the coordinates of the Tycho-2 reference stars at the epoch of the
plates. Therefore, using more accurate {\it Gaia} TGAS reference stars does not
improve astrometry. This explanation is not supported by the fact that, i) the
astrometric residuals become smaller when using only those reference stars for
which the coordinate accuracy is limited (Table~\ref{table:astrometry}), and ii)
astrometric accuracy is better when using the UCAC5 reference stars instead of
the {\it Gaia} TGAS or Tycho-2 stars.

To test whether the bias in the astrometric residuals of the {\it Gaia} TGAS
(Fig.~\ref{fig:sigma-vs-magnitude}) and HSOY
(Fig.~\ref{fig:sigma-vs-magnitude-HSOY}) data can be explained by biased proper
motion values, we add a magnitude-dependent value to the Declination proper
motion values given in the (presumably unbiased) UCAC5 catalog. The functional
form of the modification, shown in
Fig.~\ref{fig:sigma-vs-magnitude-UCAC5-mod-pm}, is made similar to the bias seen
in the astrometric residuals of the HSOY data
(Fig.~\ref{fig:sigma-vs-magnitude-HSOY}). The modification is limited to stars
with {\it Gaia} G-band magnitude greater than 12\,mag.  The amount of
modification is about 2\,mas/yr for a star with a {\it Gaia} G-band magnitude of
14\,mag.  We compute the coordinates of the UCAC5 stars at the epoch of the
plates using the modified proper motions, and then we make astrometric fits. The
residuals between the fitted and cataloged coordinates as a function of
magnitude are shown in Fig.~\ref{fig:sigma-vs-magnitude-UCAC5-mod-pm}. The
residuals along Declination are biased in a way which resembles the bias in the
proper motion values. The bias of the astrometric residuals at magnitude 14 is
about 0.15\,arcsec, which, after dividing with a time span of 100 years,
corresponds to a $\sim$1.5\,mas/yr bias in proper motion, close to the value of
the modification, $\sim$2\,mas/yr.  We thus conclude that, i) our data analysis
can detect a magnitude-dependent bias of proper motion values and give a correct
value for the bias, and ii) the bias in the astrometric residuals of the {\it
  Gaia} TGAS and HSOY data can be explained by their biased proper motion
values.

Our conclusion on the biased {\it Gaia} TGAS proper motions is supported by
\citet{Fedorov2018}.  They analyze the proper motions given in the {\it Gaia}
TGAS, HSOY, UCAC5, and PMA catalogs under the assumption that systematic
differences between the proper motions are caused by a mutual rigid-body
rotation of the reference frames of the catalogs.  They find that the published
proper motions of the Tycho-2 stars in the {\it Gaia} TGAS catalog, derived with
AGIS (Astrometric Global Iterative Solution), have a dependence on stellar
magnitude. When deriving the proper motions in a classical way (the difference
of the positions of Hipparcos/Tycho-2 -- {\it Gaia} at the corresponding epochs
divided by the span between the epochs), \citet{Fedorov2018} find that the
published and classical proper motions of Hipparcos stars in the {\it Gaia} TGAS
catalog are similar, while for Tycho-2 stars they are different. The dependence
of the proper motions of the Tycho-2 stars on stellar magnitude is eliminated
when the published {\it Gaia} TGAS proper motions are replaced with the
classical ones.

The trend seen for magnitude residuals as a function of distance from the plate
center in Figs.~\ref{fig:magresiduals} and~\ref{fig:magresiduals-3} is
contrary to what is expected for vignetting by the telescope, which attenuates
the stellar brightness outward from plate center.  The trend is probably caused
by the saturation effects of the emulsion. The stellar profiles in the plates are
larger, and thus less saturated, at larger distances from the plate center.
Therefore, we recover more of the original stellar flux at larger distances from the
plate center.  The difference between the CduC and Tycho magnitudes is zero at
about half the distance from plate center to edge (0.6\degr) because the least
squares fit of the magnitude calibration (Fig.~\ref{fig:photometry}) does not
use information about the pixel coordinates of the stars.
  
The {\it Gaia} TGAS catalog has about two million sources. The upcoming (April
2018) {\it Gaia}~DR2 is anticipated to have five-parameter astrometric solutions
for $>10^9$ sources.  Then, i) the accuracy of the astrometry is not limited by
the accuracy of the proper motion values of the reference stars, ii) we can
study the effects of the possible close companion stars that are within the
large stellar profiles of the CduC stars, and iii) {\it Gaia}~DR2 will contain
photometry not only in the G-band, but also in the blue and red colors
potentially allowing improvements in the CduC photometry calibration.

\section{Conclusions}

Six single-exposure and four triple-exposure CduC plates have been digitized
with a digital camera and a macro lens, with a resolution of
$\sim$0.7\arcsec\ per pixel.  The astrometry for the images has been derived by
using stars in the Tycho-2, {\it Gaia} TGAS (Tycho-{\it Gaia} Astrometric
Solution), UCAC5 (USNO CCD Astrograph Catalog), HSOY (Hot Stuff for One Year),
and PMA catalogs as reference stars.

The best astrometric accuracy is obtained with the UCAC5 reference stars. The
internal astrometric accuracy, that is, the deviation of the coordinates of the
stars detected within overlapping images of a single plate, is
$\sim$0.06\arcsec.  The external astrometric accuracy, that is, the deviation of the
coordinates relative to the UCAC5 reference stars at the epoch of each plate,
is $\sim$0.16\arcsec\ and $\sim$0.13\arcsec\ for the single- and triple-exposure
plates, respectively.  The plate-to-plate astrometric accuracy, that is,
the deviation of the coordinates of the stars detected within two overlapping
plates, is $\sim$0.21\arcsec\ along R.A.\ and $\sim$0.15\arcsec\ along
Declination.
The photometric accuracy is $\sim$0.28\,mag and $\sim$0.24\,mag for the single-
and triple-exposure plates, respectively.

The astrometric residuals of the UCAC5 and PMA stars at the epoch of the plates,
and thus their proper motion values, are free of a magnitude-dependent bias.
However, we believe that the R.A.\ proper motions in the {\it Gaia} TGAS catalog
and the Declination proper motions in the HSOY catalog have a
magnitude-dependent bias. The magnitude of the bias is about 2\,mas/yr at
maximum.

There is a factor of about two difference in the astrometric quality between
different plates, probably reflecting the intrinsic quality of emulsion on the
plates. The astrometric precision is limited both by the precision of the
stellar pixel positions measured on the plates and by the precision of the
proper motion values of the reference stars.

Currently, we detect magnitude equation in our plates, but the number of the
plates analyzed in our study is too low to properly determine a correction for
magnitude equation.  In the future the astrometric accuracy of our CduC plates
will be improved because we will use data from the {\it Gaia} DR2 catalog,
and we can make a correction for magnitude equation after we have digitized
more plates.

Our current method to determine and remove the distortions caused by our
instrumental setup is restricted by limitations in hardware and software, and
thus not realized in this survey. On the other hand, removal of distortions
caused by the instrumental setup would be justified only if the mechanical setup
was absolutely non-time variable, which is not the case for our current setup.
The astrometric accuracy obtained by us is equal or better than the accuracy of
the other digitizing projects of the CduC plates.

We believe that a digital camera is a better instrument than a scanner to
digitize the relatively small CduC plates. In the case of the camera Canon
EOS~5Ds, it is sufficient to take four images of a CduC plate to achieve a good
enough resolution.

\begin{acknowledgements}
We thank the late T.\ Markkanen who initiated this project, and whose
wide knowledge of the Carte du Ciel survey was essential for this project.  We
would like to thank the anonymous referee for helpful comments.  We thank
Herv\'e Bouy and Emmanuel Bertin for assistance with the usage of Astromatic
software.  This work has made use of data from the European Space Agency (ESA)
mission {\it Gaia} (\url{https://www.cosmos.esa.int/Gaia}),
processed by the {\it Gaia} Data Processing and Analysis Consortium (DPAC,
\url{https://www.cosmos.esa.int/web/Gaia/dpac/consortium}). Funding for
the DPAC has been provided by national institutions, in particular the
institutions participating in the {\it Gaia} Multilateral Agreement.
\end{acknowledgements}

\bibliographystyle{aa}
\bibliography{32662corr}

\setcounter{table}{1}

\longtab{
\begin{longtable}{c c c c c c c c c c c}
\caption{\label{table:astrometry} Results of astrometric solutions as given by
  SCAMP, both for internal and external calibration. dAXIS1 and dAXIS2 are along
  R.A.\ and Declination, respectively. The meaning of the colums: 1:
  plate number, 2: the catalog used for reference stars, 3 and 4: the internal
  RMS deviation of the astrometric fit along the axes of the images, 5: reduced
  $\chi^2$ goodness of the internal astrometric solution, 6 and 7: the external
  RMS deviation of the astrometric fit along the axes of the images, 8: reduced
  $\chi^2$ goodness of the external astrometric solution, 9: whether we use all
  the reference stars (empty) or only those stars which have positional
  uncertainty, at the epoch of the plates, less than the value given, 10: degree
  of the polynomial of the astrometric fit by SCAMP, 11: single- (1) or triple-
  (3) exposure plate. At the end of the table, the mean values of the parameters
  are given separately for the single and triple exposures. }\\
\hline\hline
Plate & Cat- & dAXIS1$_{int}$ & dAXIS2$_{int}$ & $\chi^2_{int}$   
             & dAXIS1$_{ext}$ & dAXIS2$_{ext}$ & $\chi^2_{ext}$ & Reference & Polyn. & Single/ \\  
      & alog & [arcsec]       & [arcsec]       &                        
             & [arcsec]       & [arcsec]       &                & accuracy  & degree & Triple \\    
\hline        
1 & 2 & 3 & 4 & 5 & 6 & 7 & 8 & 9 & 10 & 11\\
\hline
\endfirsthead
\caption{continued.}\\
\hline\hline            
Plate & Cat- & dAXIS1$_{int}$ & dAXIS2$_{int}$ & $\chi^2_{int}$   
             & dAXIS1$_{ext}$ & dAXIS2$_{ext}$ & $\chi^2_{ext}$ & Reference & Polyn. & Single/ \\  
      & alog & [arcsec]       & [arcsec]       &                        
             & [arcsec]       & [arcsec]       &                & accuracy  & degree & Triple \\    
\hline        
1 & 2 & 3 & 4 & 5 & 6 & 7 & 8 & 9 & 10 & 11\\
\hline
\endhead
\hline
\endfoot
 841 & {\it Gaia} TGAS  & 0.031 & 0.035  &  2.0 & 0.27 & 0.29 &   1.5 &              & 3 & 1 \\  
     & {\it Gaia} TGAS  & 0.037 & 0.036  &  2.2 & 0.22 & 0.24 &   1.6 & <0.15\arcsec & 3 & 1  \\
     & UCAC5 & 0.030 & 0.025  &  1.4 & 0.22 & 0.23 &  0.86 &              & 3 & 1 \\   
     & UCAC5 & 0.032 & 0.026  &  1.6 & 0.17 & 0.20 &  0.71 & <0.20\arcsec & 3 & 1  \\ 
     & HSOY  & 0.036 & 0.040  &  2.8 & 0.27 & 0.33 &   5.9 &              & 3 & 1 \\  
     & HSOY  & 0.034 & 0.040  &  2.6 & 0.27 & 0.32 &   5.5 & <0.27\arcsec & 3 & 1  \\ 
     & PMA   & 0.030 & 0.026  &  1.5 & 0.33 & 0.36 &  0.79 &              & 3 & 1  \\
     & PMA   & 0.029 & 0.026  &  1.5 & 0.33 & 0.37 &  0.84 & <0.32\arcsec & 3 & 1  \\
   & Tycho-2 & 0.028 & 0.028  &  1.5 & 0.24 & 0.25 &  0.85 &              & 3 & 1  \\
\hline
 844 & {\it Gaia} TGAS  & 0.032 & 0.031  &  2.0 & 0.24 & 0.19 &  0.94 &              & 2 & 1  \\ 
     & {\it Gaia} TGAS  & 0.036 & 0.034  &  2.5 & 0.21 & 0.18 &  1.1  & <0.15\arcsec & 2 & 1  \\ 
     & UCAC5 & 0.028 & 0.022  &  1.3 & 0.17 & 0.16 &  0.44 &              & 2 & 1  \\  
     & UCAC5 & 0.028 & 0.022  &  1.3 & 0.15 & 0.14 &  0.38 & <0.20\arcsec & 2 & 1  \\ 
     & HSOY  & 0.030 & 0.026  &  1.6 & 0.27 & 0.26 &   2.2 &              & 2 & 1  \\ 
     & HSOY  & 0.030 & 0.026  &  1.6 & 0.27 & 0.26 &   2.1 & <0.27\arcsec & 2 & 1  \\  
     & PMA   & 0.028 & 0.021  &  1.2 & 0.28 & 0.29 &  0.63 &              & 2 & 1 \\
     & PMA   & 0.028 & 0.020  &  1.2 & 0.28 & 0.28 &  0.63 & <0.32\arcsec & 2 & 1 \\
   & Tycho-2 & 0.029 & 0.024  &  1.4 & 0.19 & 0.19 &  0.55 &              & 2 & 1  \\ 
\hline
 883 & {\it Gaia} TGAS  & 0.048 & 0.041 & 3.8 & 0.19 & 0.15 & 1.2  &               & 2 & 1  \\ 
     & {\it Gaia} TGAS  & 0.047 & 0.041 & 3.6 & 0.16 & 0.13 & 1.1  &  <0.15\arcsec & 2 & 1  \\ 
     & UCAC5 & 0.040 & 0.038 & 2.7 & 0.19 & 0.16 & 0.63 &               & 2 & 1  \\   
     & UCAC5 & 0.039 & 0.036 & 2.6 & 0.16 & 0.14 & 0.51 & <0.20\arcsec & 2 & 1  \\  
     & HSOY  & 0.042 & 0.041 & 3.0 & 0.25 & 0.27 & 2.7  &               & 2 & 1  \\ 
     & HSOY  & 0.042 & 0.041 & 3.0 & 0.24 & 0.27 & 2.3  & <0.27\arcsec & 2 & 1  \\ 
     & PMA   & 0.039 & 0.035 & 2.4 & 0.26 & 0.28 & 0.55 &              & 2 & 1 \\
     & PMA   & 0.038 & 0.033 & 2.4 & 0.26 & 0.29 & 0.61 & <0.32\arcsec & 2 & 1 \\  
   & Tycho-2 & 0.042 & 0.039 & 3.0 & 0.19 & 0.17 & 0.53 &               & 2 & 1  \\
\hline
 886 & {\it Gaia} TGAS  &  0.033 & 0.027 &   1.8 & 0.25 & 0.19 &   1.4 &              & 2 & 1  \\  
     & {\it Gaia} TGAS  &  0.035 & 0.027 &   1.9 & 0.21 & 0.16 &   1.0 & <0.15\arcsec & 2 & 1  \\ 
     & UCAC5 &  0.027 & 0.022 &   1.1 & 0.18 & 0.17 &  0.56 &              & 2 & 1  \\    
     & UCAC5 &  0.027 & 0.022 &   1.2 & 0.14 & 0.14 &  0.40 & <0.20\arcsec & 2 & 1  \\ 
     & HSOY  &  0.027 & 0.022 &   1.2 & 0.40 & 0.45 &   3.6 &              & 2 & 1  \\  
     & HSOY  &  0.028 & 0.022 &   1.2 & 0.39 & 0.44 &   3.6 & <0.27\arcsec & 2 & 1  \\  
     & PMA   &  0.026 & 0.020 &   1.1 & 0.30 & 0.30 &  0.60 &              & 2 & 1 \\
     & PMA   &  0.026 & 0.021 &   1.1 & 0.30 & 0.31 &  0.66 & <0.32\arcsec & 2 & 1 \\    
   & Tycho-2 &  0.026 & 0.023 &   1.2 & 0.22 & 0.19 &  0.59 &              & 2 & 1  \\
\hline
 890 & {\it Gaia} TGAS  & 0.079 & 0.065 &  9.0 & 0.36 & 0.32 &   4.4 &               & 3 & 1  \\  
     & {\it Gaia} TGAS  & 0.083 & 0.068 &  9.9 & 0.36 & 0.30 &   4.8 & <0.15\arcsec  & 3 & 1  \\  
     & UCAC5 & 0.057 & 0.047 &  4.7 & 0.33 & 0.31 &   2.1 &               & 3 & 1  \\     
     & UCAC5 & 0.055 & 0.044 &  4.3 & 0.31 & 0.30 &   1.9 &  <0.20\arcsec & 3 & 1  \\ 
     & HSOY  & 0.065 & 0.051 &  6.0 & 0.43 & 0.41 &   7.3 &               & 3 & 1  \\
     & HSOY  & 0.065 & 0.051 &  6.0 & 0.42 & 0.40 &   7.2 &  <0.27\arcsec & 3 & 1  \\
     & PMA   & 0.067 & 0.064 &  8.0 & 0.40 & 0.39 &   1.0 &               & 3 & 1  \\
     & PMA   & 0.071 & 0.082 & 10.0 & 0.40 & 0.39 &   1.2 &  <0.32\arcsec & 3 & 1  \\
   & Tycho-2 & 0.052 & 0.039 &  3.6 & 0.37 & 0.34 &   1.7 &               & 3 & 1  \\  
\hline   
 892 & {\it Gaia} TGAS  & 0.037 & 0.032 &  2.2 & 0.23 & 0.18 &   1.3 &              & 2 & 1  \\  
     & {\it Gaia} TGAS  & 0.039 & 0.032 &  2.4 & 0.22 & 0.17 &   1.5 & <0.15\arcsec & 2 & 1  \\ 
     & UCAC5 & 0.035 & 0.025 &  1.8 & 0.17 & 0.15 &  0.64 &              & 2 & 1  \\  
     & UCAC5 & 0.032 & 0.024 &  1.6 & 0.16 & 0.14 &  0.56 & <0.20\arcsec & 2 & 1  \\    
     & HSOY  & 0.044 & 0.031 &  2.7 & 0.28 & 0.27 &  4.3  &              & 2 & 1  \\ 
     & HSOY  & 0.043 & 0.030 &  2.6 & 0.28 & 0.26 &  4.2  & <0.27\arcsec & 2 & 1  \\ 
     & PMA   & 0.026 & 0.024 &  1.2 & 0.30 & 0.30 &  0.70 &              & 2 & 1  \\
     & PMA   & 0.027 & 0.025 &  1.3 & 0.29 & 0.29 &  0.71 & <0.32\arcsec & 2 & 1  \\
   & Tycho-2 & 0.033 & 0.027 &  1.8 & 0.23 & 0.21 &  0.77 &              & 2 & 1  \\           
\hline
 854 & {\it Gaia} TGAS  & 0.019 &  0.014  &   1.2 &  0.20 &  0.14  &  0.58 &              & 2 & 3  \\
     & {\it Gaia} TGAS  & 0.020 &  0.016  &   1.2 &  0.17 &  0.13  &  0.63 & <0.15\arcsec & 2 & 3  \\ 
     & UCAC5 & 0.024 &  0.023  &   1.0 &  0.15 &  0.14  &  0.40 &              & 2 & 3  \\     
     & UCAC5 & 0.024 &  0.022  &   1.0 &  0.14 &  0.13  &  0.35 & <0.20\arcsec & 2 & 3  \\   
     & HSOY  & 0.025 &  0.025  &   1.2 &  0.22 &  0.23  &  1.5  &              & 2 & 3  \\ 
     & HSOY  & 0.026 &  0.026  &   1.2 &  0.22 &  0.23  &  1.5  & <0.27\arcsec & 2 & 3  \\  
     & PMA   & 0.029 &  0.023  &   1.4 &  0.28 &  0.30  &  0.61 &              & 2 & 3 \\
     & PMA   & 0.024 &  0.024  &   1.1 &  0.26 &  0.28  &  0.55 & <0.32\arcsec & 2 & 3 \\  
   & Tycho-2 & 0.018 &  0.016  &   1.1 &  0.16 &  0.17  &  0.48 &              & 2 & 3  \\    
\hline
 887 & {\it Gaia} TGAS  & 0.021 & 0.025   &  1.9 & 0.19 &  0.21  &   1.5 &              & 2 & 3  \\    
     & {\it Gaia} TGAS  & 0.022 & 0.026   &  2.0 & 0.18 &  0.20  &   1.6 & <0.15\arcsec & 2 & 3  \\ 
     & UCAC5 & 0.028 & 0.026   &  0.9 & 0.17 &  0.20  &  0.67 &              & 2 & 3  \\   
     & UCAC5 & 0.028 & 0.026   &  0.9 & 0.15 &  0.17  &  0.55 & <0.20\arcsec & 2 & 3  \\  
     & HSOY  & 0.035 & 0.034   &  1.9 & 0.22 &  0.31  &   3.3 &              & 2 & 3  \\
     & HSOY  & 0.034 & 0.035   &  1.9 & 0.22 &  0.31  &   3.3 & <0.27\arcsec & 2 & 3  \\  
     & PMA   & 0.066 & 0.059   & 7.6  & 0.33 &  0.33  & 0.59  &              & 2 & 3 \\
     & PMA   & 0.058 & 0.112   & 20.0 & 0.31 &  0.36  & 0.80  & <0.32\arcsec & 2 & 3 \\ 
   & Tycho-2 & 0.019 & 0.019   &  1.0 & 0.18 &  0.22  &  0.65 &              & 2 & 3  \\    
\hline
 894 & {\it Gaia} TGAS  & 0.028 & 0.035   &  2.8 & 0.16 &  0.24  &   1.9 &              & 2 & 3  \\   
     & {\it Gaia} TGAS  & 0.030 & 0.036   &  3.1 & 0.15 &  0.21  &   1.9 & <0.15\arcsec & 2 & 3  \\ 
     & UCAC5 & 0.032 & 0.027   &  1.1 & 0.18 &  0.22  &  0.67 &              & 2 & 3  \\    
     & UCAC5 & 0.032 & 0.027   &  1.1 & 0.16 &  0.19  &  0.59 & <0.20\arcsec & 2 & 3  \\   
     & HSOY  & 0.037 & 0.035   &  1.8 & 0.32 &  0.38  &   3.4 &              & 2 & 3  \\
     & HSOY  & 0.037 & 0.036   &  1.8 & 0.32 &  0.37  &   3.4 & <0.27\arcsec & 2 & 3  \\  
     & PMA   & 0.031 & 0.026   & 0.98 & 0.29 &  0.34  &  0.65 &              & 2 & 3 \\
     & PMA   & 0.032 & 0.027   &  1.1 & 0.30 &  0.34  &  0.69 & <0.32\arcsec & 2 & 3 \\
   & Tycho-2 & 0.022 & 0.018   &  1.2 & 0.16 &  0.21  &  0.61 &              & 2 & 3  \\  
\hline
 896 & {\it Gaia} TGAS  & 0.020 & 0.019  &  0.75 & 0.16 &  0.13  &   0.5 &              & 2 & 3  \\    
     & {\it Gaia} TGAS  & 0.020 & 0.019  &  0.79 & 0.13 &  0.12  &   0.5 & <0.15\arcsec & 2 & 3  \\  
     & UCAC5 & 0.023 & 0.021  &  0.71 & 0.16 &  0.15  &  0.55 &              & 2 & 3  \\     
     & UCAC5 & 0.023 & 0.021  &  0.73 & 0.14 &  0.13  &  0.46 & <0.20\arcsec & 2 & 3  \\  
     & HSOY  & 0.026 & 0.022  &  0.84 & 0.21 &  0.20  &   1.5 &              & 2 & 3  \\ 
     & HSOY  & 0.026 & 0.022  &  0.84 & 0.21 &  0.20  &   1.5 & <0.27\arcsec & 2 & 3  \\ 
     & PMA   & 0.024 & 0.022  &  0.77 & 0.23 &  0.25  &  0.47 &              & 2 & 3 \\
     & PMA   & 0.025 & 0.022  &  0.78 & 0.24 &  0.24  &  0.49 & <0.32\arcsec & 2 & 3 \\
   & Tycho-2 & 0.021 & 0.019  &  0.82 & 0.19 &  0.18  &  0.51 &              & 2 & 3  \\
\hline
\hline
Mean   & {\it Gaia} TGAS  & 0.043 &   0.039   &   3.5  &   0.26  &   0.22   &   1.8 &              &   & 1  \\  
values & {\it Gaia} TGAS  & 0.046 &   0.040   &   3.8  &   0.23  &   0.20   &   1.9 & <0.15\arcsec &   & 1  \\
       & UCAC5 & 0.036 &   0.030   &   2.2  &   0.21  &   0.20   &  0.87 &              &   & 1  \\
       & UCAC5 & 0.035 &   0.029   &   2.1  &   0.18  &   0.18   &  0.74 & <0.20\arcsec &   & 1  \\ 
       & HSOY  & 0.040 &   0.035   &   2.9  &   0.32  &   0.33   &   4.3 &              &   & 1  \\ 
       & HSOY  & 0.040 &   0.035   &   2.8  &   0.31  &   0.33   &   4.1 & <0.27\arcsec &   & 1  \\ 
       & PMA   & 0.036 &   0.032   &   2.6  &   0.31  &   0.32   &  0.71 &              &   & 1  \\
       & PMA   & 0.036 &   0.035   &   2.9  &   0.31  &   0.32   &  0.78 & <0.32\arcsec &   & 1  \\  
       & Tycho-2 & 0.035 &   0.030   &   2.1  &   0.24  &   0.23   &  0.83 &            &   & 1  \\
\hline
Mean   & {\it Gaia} TGAS  & 0.022  &  0.023  &  1.7  &  0.18  &  0.18  &   1.1 &              &   & 3  \\ 
values & {\it Gaia} TGAS  & 0.023  &  0.024  &  1.8  &  0.16  &  0.17  &   1.2 & <0.15\arcsec &   & 3  \\
       & UCAC5 & 0.027  &  0.024  &  0.93  &  0.16  &  0.18  &  0.57 &              &   & 3  \\  
       & UCAC5 & 0.027  &  0.024  &  0.94  &  0.15  &  0.16  &  0.49 & <0.20\arcsec &   & 3  \\  
       & HSOY  & 0.031  &  0.029  &   1.4  &  0.24  &  0.28  &   2.4 &              &   & 3  \\  
       & HSOY  & 0.031  &  0.029  &   1.4  &  0.24  &  0.28  &   2.4 & <0.27\arcsec &   & 3  \\ 
       & PMA   & 0.037  &  0.032  &   2.7  &  0.28  &  0.31  &  0.58 &              &   & 3  \\
       & PMA   & 0.035  &  0.046  &   5.7  &  0.28  &  0.31  &  0.63 & <0.32\arcsec &   & 3 \\  
       & Tycho-2 & 0.020  &  0.018   &   1.0  &   0.17  &   0.20  &   0.56 &            &   & 3  \\
\end{longtable}
}

\end{document}